\documentclass[prd,reprint,no title page,twoside,onecolumn,showpacs,
               nofootinbib, longbibliography,amsmath,amssymb,showkeys,
               floatfix]{revtex4-1}

\usepackage{mathtools}
\usepackage{graphicx}   
\usepackage{dcolumn}    
\usepackage{bm}         
\usepackage{multirow}   
\usepackage{bigints}    

\providecommand*{\defeq}{\mathrel{\vcenter            %
                        {\baselineskip0.5ex           %
                         \lineskiplimit0pt            %
                         \hbox{\scriptsize.}          
                         \hbox{\scriptsize.}}} =}     
\providecommand*{\eqdef}{= \mathrel{\vcenter          
                        {\baselineskip0.5ex           
                         \lineskiplimit0pt            %
                         \hbox{\scriptsize.}          %
                         \hbox{\scriptsize.}}}}       %

\begin{document}

\title{COSMOLOGY WITHOUT EINSTEIN'S ASSUMPTION THAT INERTIAL MASS PRODUCES
GRAVITY}

\author{HOMER G. ELLIS}

\affiliation{Department of Mathematics, University Colorado Boulder, 395 UCB,\\
Boulder, Colorado 80309-0395, United States of America \\
ellis@euclid.colorado.edu}

\date{May 2, 2015}

\begin{abstract}
Giving up Einstein's assumption, implicit in his 1916 field equations, that
inertial mass, even in its appearance as energy, is equivalent to active
gravitational mass and therefore is a source of gravity allows revising the
field equations to a form in which a positive cosmological constant is seen to
(mis)represent a uniform negative net mass density of gravitationally
attractive and gravitationally repulsive matter.  Field equations with both
positive and negative active gravitational mass densities of both primordial
and continuously created matter incorporated, along with two scalar fields to
`relax the constraints' on the space-time geometry, yield cosmological
solutions that exhibit inflation, deceleration, coasting, acceleration, and a
`big bounce' instead of a `big bang', and provide good fits to a Hubble diagram
of type~Ia supernovae data.  The repulsive matter is identified as the back
sides of the `drainholes' introduced by the author in 1973 as solutions of
those same field equations.  Drainholes (prototypical examples of `traversable
wormholes') are topological tunnels in space which gravitationally attract on
their front, entrance sides and repel more strongly on their back, exit sides.
The front sides serve both as the gravitating cores of the visible, baryonic
particles of primordial matter and as the continuously created, invisible
particles of the `dark matter' needed to hold together the large scale
structures seen in the universe; the back sides serve as the misnamed `dark
energy' driving the current acceleration of the expansion of the universe.
Formation of cosmic voids, walls, filaments, and nodes is attributed to
expulsion of drainhole entrances from regions populated by drainhole exits, and
accumulation of the entrances on boundaries separating those regions.
\end{abstract}



\maketitle

\vskip -30pt

\tableofcontents

\vskip 2\baselineskip
\hspace{0.01\textwidth}
\parbox{0.88\textwidth}{
{\textit{\textbf{
It can scarcely be denied that the supreme goal of all theory is to make the
irreducible basic elements as simple and as few as possible without having to
surrender the adequate representation of a single datum of experience.
}}} \hfill Albert Einstein}

\vskip 2\baselineskip
\hspace{0.01\textwidth}
\parbox{0.88\textwidth}{
{\sl
The beginning of wisdom is to call things by their proper name.
}
\flushright Chinese Proverb (after Confucius)}

\section{Introduction}
\label{sec1}

In an article published in Journal of Mathematical Physics in 1973 I derived
and analyzed in detail a model for a gravitating particle that was an
improvement on the Schwarzschild blackhole model~\cite{elli1}.  This static and
spherically symmetric space-time manifold, discovered independently at about
the same time by K.~A.~Bronnikov~\cite{bron} and formally some years earlier by
O.~Bergmann and R.~Leipnik who rejected it for
``physical reasons''~\!\!\cite{bele}, I termed a `drainhole' with `ether'
flowing through it.  The manifold is geodesically complete, singularity-free,
and devoid of horizons.  It comprises a topological hole (the `drainhole')
connecting two spatial regions, an `upper' and a `lower', on which there is a
space-time vector field (the `ether-flow' vector field) representing the
velocities of test particles free-falling from rest at infinity in the upper
region, into and through the hole, and out into the lower region, accelerating
downward all the way.  The upper region being asymptotic to a Schwarzschild
manifold with positive mass parameter $m$, the lower region is asymptotic to a
Schwarzschild manifold with negative mass parameter
$\bar m = -m e^{m \pi/\sqrt{n^2 - m^2}}$, where $n$ is a parameter that
determines the size of the hole.  Thus the drainhole attracts test particles on
its high side, but repels them on the low side more strongly, in the ratio
$-\bar m/m = e^{m \pi/\sqrt{n^2 - m^2}} > 1$.  The drainhole can be thought of
as a kind of natural accelerator of the `gravitational ether', drawing it in on
the high side and expelling it more forcefully on the low side.  To avoid
ambiguities associated with the term `ether' one could say it is space itself
that is flowing into and through the drainhole, a substitution that would accord
well with Einstein's insight that the concepts of space and of a gravitational
ether are essentially interchangeable~\cite{eins1}.

In the 1973 paper I wrote that a ``speculative extrapolation from the asymmetry
between $m$ and $\bar m$ is that the universe expands because it contains more
negative mass than positive, each half-particle of positive mass $m$ being
slightly overbalanced by a half-particle of negative mass $\bar m$ such that
$-\bar m > m$.''  This speculation lay dormant until the beginning of 2006, when
it occurred to me that the same mechanism might be used to explain not only the
expansion of the universe but also the recently discovered acceleration of that
expansion.  To properly exploit that idea I have found it necessary to reject
three of the assumptions that have been built into standard relativistic
cosmological theory from its earliest days.  The first is Einstein's implicit
assumption that active gravitational mass and passive-inertial mass are the same
thing, consequently that passive-inertial mass is a source of gravity.  The
second, which uncritical acceptance of the first gives rise to, is that if a
field thought to be associated with some form of matter couples to geometry in
the field equations of space-time with the `wrong' polarity (the `wrong sign' of
the coupling constant), then that form of matter is a `phantom' or `ghost' form
that can exist only in `exotic' circumstances.  The third is that every scalar
field included in a variational principle of relativistic gravitational theory
must represent some form of matter and must therefore have its own separate
field equation produced by variation of that field in the action integral of
the variational principle.

As I shall show, denial of these three assumptions is logically consistent and
allows one to arrive at a purely geometric theory of gravitation that produces
a singularity-free cosmological model of the universe that fits with very good
precision Hubble diagram data from observations of type~Ia supernovae.
This model features a `big bounce' (instead of a `big bang'), rapid inflation
out of the bounce, and a `graceful exit' from the inflation into a long period
of decelerative coasting, followed by a transition to an ultimately
exponential, de Sitter-like accelerating expansion.  In addition, attributing
the accelerating expansion to the existence of drainholes provides explanations
for dark matter, dark `energy', and the formation of cosmic voids, walls,
filaments, and nodes.  The model is offered as a replacement for and
improvement on the standard $\Lambda$CDM cosmological model.

\section{Einstein's Implicit Assumption}
\label{sec2}

In his 1916 paper Die Grundlage der allgemeinen
Relativit\"atstheorie~\cite{eins2} that gave a thorough presentation of the
theory of gravity he had worked out over the preceding decade, Albert Einstein
made implicitly an assumption that does not hold up well under scrutiny.
Stripped down to its barest form the assumption is that inertial mass, and by
extension energy {\it via} $E = m c^2$, is a source of gravity and must
therefore be coupled to the gravitational potential in the field equations of
the general theory of relativity.  The train of thought that brought him to
this conclusion is described in \S 16, where he sought to extend his field
equations for the vacuum,
${\bm R}_{\alpha \beta} - \frac12 {\bm R} g_{\alpha \beta} = 0$ as
currently formulated, to include the contribution of a continuous distribution
of gravitating matter of density~$\rho$, in analogy to the extension of the
Laplace equation $\nabla^2 \phi = 0$ for the newtonian gravitational potential
$\phi$ to the Poisson equation $\nabla^2 \phi = 4 \pi \kappa \rho$, where
$\kappa$ is Newton's gravitational constant.  Einstein referred to $\rho$ as
the ``density of matter'', without specifying what was meant by `matter' or its
`density'.  Invoking the special theory's identification of ``inert mass'' with
``energy, which finds its complete mathematical expression in\,\ldots the
energy-tensor'', he concluded that ``we must introduce a corresponding
energy-tensor of matter ${\text T}^\alpha_\sigma$\,''.  Further describing this
energy-tensor as ``corresponding to the density~$\rho$ in Poisson's equation'',
he arrived in \S 19 at the extended field equations
{${\bm R}_{\alpha \beta} - \frac12 {\bm R} g_{\alpha \beta}
= \frac{8 \pi \kappa}{c^2} T_{\alpha \beta}$}, in which, for a ``frictionless
adiabatic fluid'' of ``density''~$\rho$, ``pressure''~$p$ (a form of kinetic
energy), and proper four-velocity distribution~$u^\alpha$, he took
$T^{\alpha \beta}$ to be $\rho u^\alpha u^\beta - p g^{\alpha \beta}$
(revised to $(\rho + p/c^2)u^\alpha u^\beta - p g^{\alpha \beta}$ by his
redefinition $\rho \to \rho + p/c^2$).

Clearly, Einstein's procedure fails to distinguish between the `active
gravitational mass' of matter, which measures how much gravity it produces and
is the sole contributor to the ``density of matter'' in Poisson's equation, and
the ``inert mass'' of matter, which measures how much it accelerates in
response to forces applied to it, an effect entirely different from the
production of gravity.  These two conceptually different masses, along with yet
a third, all occur in Newton's gravitational equation
\begin{equation}
m_{\text{inertial}} \, a_{\text B}
 = F_{\text{AB}}
 = -\kappa \frac{m_{\text{passive}} M_{\text{active}}}{r^2} \, ,
\label{eqn1}
\end{equation}
in which $M_{\text{active}}$ is the active gravitational mass of a gravitating
body A, $m_{\text{inertial}}$ is the inertial (``inert'') mass of a body B whose
acceleration $a_{\text B}$ is caused by the gravity produced by~A, and
$m_{\text{passive}}$ is the passive gravitational mass of B, a measure of the
strength of B's `sensing' of the gravitational field around~A.  That in suitable
units $m_{\text{inertial}} = m_{\text{passive}}$ for all bodies is another way
of saying that all bodies respond with the same accelerations to the same
gravitational fields, that, in consequence, the notion of a gravitational field
is more fundamental than the notion of a gravitational force.  Simple thought
experiments of Galileo (large stone and smaller stone tied
together)~\cite{gali} and Einstein (body suspended by a rope in an
elevator)~\cite{eins3} make it clear that bodies do all respond alike --- an
observation now treated as a principle, the (weak) `principle of equivalence',
experimentally, if somewhat redundantly, well confirmed.

That this passive-inertial mass
$m_{\text{passive-inertial}}$ ($= m_{\text{passive}} = m_{\text{inertial}}$) has
any relation to active gravitational mass is not apparent in Eq.~(\ref{eqn1}),
where, unlike $m_{\text{inertial}}$ and $m_{\text{passive}}$,
$M_{\text{active}}$ quantifies a property of A, not of B.  But Newton's
equation for the gravitational action of B on A reads
\begin{equation}
M_{\text{inertial}} \, a_{\text A}
 = F_{\text{BA}}
 = -\kappa \frac{M_{\text{passive}} m_{\text{active}}}{r^2} \, .
\label{eqn2}
\end{equation}
Application of Newton's law of action and reaction allows the inference that
$F_{\text{AB}}$ and $F_{\text{BA}}$ have the same magnitude, from which
follows that
$m_{\text{active}}/m_{\text{passive}} = M_{\text{active}}/M_{\text{passive}}$,
hence that the ratio of active gravitational mass to passive gravitational
mass, thus to inertial mass, is the same for all bodies.  It would seem likely
that Einstein relied, either consciously or unconsciously, on this consequence
of Newton's laws when he assumed that ``inert mass'' should contribute to the
``density of matter'' as a source of gravity in the field equations.

Newton's law of action and reaction is applicable to the bodies A and B only
under the condition that gravity acts at a distance instantaneously, that is,
at infinite propagation speed.  But the general theory of relativity
Einstein was expounding is a field theory in which gravitational effects
propagate at finite speed.  Within his own theory of gravity there is,
therefore, no obvious justification for Einstein's assumption that inertial
mass (and therefore energy) is equivalent to active gravitational mass.
Indeed, in Newton's and in Einstein's theory of gravity in the vacuum the
concept of inertial mass is irrelevant, if not in fact nonexistent.  In
Newton's case cancellation of $m_{\text{inertial}}$ with $m_{\text{passive}}$
from Eq.~(\ref{eqn1}) leaves only $M_{\text{active}}$ to determine the
gravitational influence of A on B, namely, an acceleration independent of any
property of B.  In Einstein's case the same equivalence principle cancellation
is effected by his stipulation that test particles in empty space follow the
geodesics of the gravitational field as described by the space-time geometry
there.  That this stipulation leaves in place only the active gravitational
mass concept is amply illustrated by the Schwarzschild solution
$(1 - 2M/r) c^2 dT^2 - (1 - 2M/r)^{-1} dr^2 - r^2 d\Omega^2$ of the vacuum field
equations ${\bm R}_{\alpha \beta} - \frac12 {\bm R} g_{\alpha \beta} = 0$,
in which the only mass parameter to appear is the $M$ that determines the
curvatures of the space-time metric, thus the strength of the gravitational
field of the point particle located at $r = 0$.

Einstein's not recognizing that the inference of equivalence between the
concepts of inertial mass and active mass was based on a false premise, then
using that inferred equivalence as justification for extending his vacuum field
equations to the nonvacuum case in the way he did, was a fundamental error.  Few
would fault him for such an error made in the midst of the intense intellectual
endeavor to construct his general theory of relativity.  One can speculate that
had he avoided the error he might have discovered in the years after 1916 a
number of important things that others discovered only many years
later~\cite{elli2}.


The failure of Einstein's attempt to introduce inertial mass into his theory by
way of his energy-tensor of matter does not preclude the possibility that the
concept can be introduced there by some other means.  There are in fact good
reasons to believe that it should be introduced (or discovered) there as a
facet of the space-time geometry, related in some way to the active mass
concept.  There is, for example, the seemingly universal coincidence that, in
physics beyond the merely gravitational, wherever there is matter made of atoms
there are to be found both inertial mass and active gravitational mass.
Indeed, the fact that Newton's theory gives results that describe as well as
they do the motions of the planets and their satellites would argue for some
proportionality between $m_{\text{active}}$ and $m_{\text{passive}}$ for such
matter {\it in bulk} --- not, however, for each individual constituent of such
matter.  A 1986 analysis of lunar ranging data concluded that the ratio of
$m_{\text{active}}$ to $m_{\text{passive}}$ for aluminum differs from that for
iron by less than $4 \times 10^{-12}$~\cite{babu}.  An earlier,
Cavendish-balance experiment had put a limit of $5 \times 10^{-5}$ on the
difference of these ratios for bromine and fluorine~\cite{kreu}.  But those
results are {\it only} for matter in bulk, that is, matter made of atoms and
molecules.  It is entirely possible that electrons, for example, do not
gravitate at all, for no one has ever established by direct observation that
they do, nor is it likely that anyone will soon do so.  There is in the
literature an argument that purports to show that if the ratio
$m_{\text{active}}/m_{\text{passive}}$ is the same for two species of bulk
matter, then electrons must be generators of gravity~\cite{ungi}, but that
argument can be seen on careful examination to rest on an unrecognized, hidden
assumption of its own, namely that, in simplest form, the gravitational field
of a hydrogen atom at a distance could be distinguished from that of a neutron
at the same distance~\cite{elli3} --- another assumption no one has tested, or
is likely soon to test, by direct observation.

Einstein's assumption that energy and inertial mass are sources of gravity has
survived to the present virtually unchallenged.\footnote{Curiously, Herman
Bondi in a paper in 1957 carefully distinguished between passive-inertial mass
and active mass, then in the same paper adopted Einstein's ``energy-tensor''
which ignores the distinction~\cite{bond}.} It has generated a number of
consequences that have directed much of the subsequent research in gravitation
theory --- indeed, misdirected it if his assumption is wrong.  Among them are
these:

\begin{itemize}

\item The impossibility, according to Penrose--Hawking singularity theorems,
of avoiding singularities in the geometry of space-time without invoking
`negative energy', which is really just energy coupled to gravity with polarity
(`sign') opposite to that of the coupling of matter to gravity.

\item The presumption that the extra, fifth  dimension in Kaluza--Klein theory
must be a spatial dimension rather than a dimension of another type.

\item The belief that all the extra dimensions in higher-dimensional theories
must be spatial, causing the expenditure of much effort in mental gymnastics to
explain why they are not apparent to our senses in the way that the familiar
three spatial dimensions are.

\end{itemize}

\noindent
Denying Einstein's assumption relieves one of the burden of these troublesome
conclusions and opens the door to other, more realistic ones.

\enlargethispage{2.5\baselineskip}
\section{New, Improved Field Equations}
\label{sec3}

If Einstein's assumption is to be disallowed, then his source tensor for a
continuous distribution of gravitating matter,
$T^{\alpha \beta} = (\rho \, + \, p/c^2) u^\alpha u^\beta - p g^{\alpha \beta}$,
must be modified or replaced.  One might think simply to drop the pressure
terms and take $T^{\alpha \beta} = \rho u^\alpha u^\beta$, the energy-momentum
tensor of the matter.  This would be inconsistent, for the $\rho$ in that
tensor is the density of inertial mass, which we are now not assuming to be the
same as active gravitational mass.  What should one do instead?

At the same time that Einstein was creating his field equations David Hilbert
was deriving the field equations for (in particular) empty space from the
variational principle
$\delta \! \bigintsss \!\! {\bm R}\,|g|^{\frac12} d^4\!x = 0$~\cite{hilb}.  This
is the most straightforward extension to the general relativity setting of the
variational principle $\delta \! \bigintsss \! |\nabla \phi|^2 \, d^3\!x = 0$,
whose Euler--Lagrange equation is equivalent to the Laplace equation
$\nabla^2 \phi = 0$ for the newtonian potential $\phi$.  Modifying that
principle to
$\delta \! \bigintsss \! (|\nabla \phi|^2 + 8 \pi \kappa \mu \phi) \, d^3\!x
= 0$,
where $\mu$ is the density of the {\it active\/} gravitational mass of matter,
generates the Poisson equation $\nabla^2 \phi = 4 \pi \kappa \mu$.  The most
straightforward extension of this modified principle to general relativity is
\begin{equation}
\delta \! \int ({\bm R} - \textstyle\frac{8 \pi \kappa}{c^2} \mu)
               \, |g|^{\frac12} d^4\!x = 0 \, ,
\label{eqn3}
\end{equation}
for which the Euler--Lagrange equations are equivalent to
\begin{equation}
{\bm R}_{\alpha \beta} - \textstyle{\frac12} {\bm R} \, g_{\alpha \beta}
 = -\frac{4 \pi \kappa}{c^2} \mu g_{\alpha \beta} \, ,
\label{eqn4}
\end{equation}
\noindent
which makes $T_{\alpha \beta} = -\frac12 \mu g_{\alpha \beta}$.  Equivalent to
this equation is
${\bm R}_{\alpha \beta} = \frac{4 \pi \kappa}{c^2} \mu g_{\alpha \beta}$, the
00 component of which reduces in the slowly varying, weak field approximation
to the Poisson equation, with $\phi = \frac12 (g_{00} - c^2)$.

The vanishing of the divergence of the Einstein tensor field
${\bm R}_{\alpha \beta} - \textstyle{\frac12} {\bm R} \, g_{\alpha \beta}$
in Eq.~(\ref{eqn4}) entails that
$0 = {{T_{\alpha}}^{\beta}}_{:\beta}
= -\frac12 (\mu_{.\beta}{g_\alpha}^\beta + \mu {{g_{\alpha}}^{\beta}}_{:\beta})
= -\frac12 \mu_{.\alpha}$, hence that $\mu$ is both spatially and temporally
constant.  These constraints on $\mu$ being unduly restrictive, further
modification is in order.

To widen the range of space-time geometries admitted by the field equations one
can in the usual way add to the action integrand of Eq.~(\ref{eqn3}) terms
related to such things as scalar fields and electromagnetic fields.  In
particular, one can add a cosmological constant term, changing the integrand to
${\bm R} - \textstyle\frac{8 \pi \kappa}{c^2} \mu + 2 \Lambda$ and the field
equations to
\begin{equation}
{\bm R}_{\alpha \beta} - \textstyle{\frac12} {\bm R} \, g_{\alpha \beta}
 = -\frac{4 \pi \kappa}{c^2} \mu \, g_{\alpha \beta}
    + \Lambda  \, g_{\alpha \beta} 
 = -\frac{4 \pi \kappa}{c^2} (\mu + \bar \mu) \, g_{\alpha \beta} \, ,
\label{eqn5}
\end{equation}
where $\bar \mu = -\frac{c^2}{4 \pi \kappa} \Lambda$.  A positive cosmological
constant $\Lambda$ thus appears in this context to be a (mis)representation of
a negative active mass density $\bar \mu$ of a continuous distribution of
gravitationally repulsive matter.  The same field equations are obtained by
changing the integrand to ${\bm R} + 2 \Lambda$ and setting
$\Lambda =  -\frac{4 \pi \kappa}{c^2} (\mu + \bar \mu)$, thus associating a
positive $\Lambda$ with a negative {\it net} active mass density of gravitating
matter, some attractive, some repulsive.  As suggested in the Introduction, an
excess of the negative active mass density $\bar \mu$ of repulsive matter over
the positive density $\mu$ of attractive matter could drive an accelerating
cosmic expansion (and in the process would solve the vexing `Cosmological
Constant Problem' by identifying $-\frac{c^2}{4 \pi \kappa} \Lambda$ as the net
density $\mu + \bar \mu$ of gravitating matter).  Leaving for later a full
discussion of drainholes as the source of such a density imbalance
(cf. Sec.~\ref{sec6.1}), we can explore the consequences of an imbalance by
studying cosmological solutions of field equations that incorporate a positive
mass density $\mu$, a negative mass density $\bar \mu$, and scalar fields $\phi$
(not the newtonian $\phi$) and $\psi$, all folded into the space-time geometry
{\it via\/} the variational principle
\begin{equation}
\delta \! \int [{\bm R}
- \textstyle{\frac{8 \pi \kappa}{c^2}} (\mu + \bar \mu)
+ 2 \, \phi^{.\gamma} \phi_{.\gamma}
- 2 \, \psi^{.\gamma} \psi_{.\gamma}] \, |g|^{\frac12} d^4\!x = 0 \, .
\label{eqn6}
\end{equation}

In deriving field equations from this variational principle I will vary only
the space-time metric.  Not varying the densities is normal, but not varying
the scalar fields goes against orthodox practice.  The rationale for leaving
them unvaried is this: In a space-time manifold the geometry is determined by
the metric alone, so only the metric should participate in the extremizing of
the action; to vary the scalar fields would be to treat them as something
extraneous to the metric thus to the geometry, whereas their proper role should
be simply to introduce a useful flexibility into the extremizing process,
{\it not\/} to represent explicit contributions to gravity by the `energy'
fields of distributions of scalar matter.  Varying neither the scalar fields
nor the densities, but only the metric, is in keeping with Einstein's guiding
principle that geometry should be able to explain all of physics.

Looked at another way, including the terms
$-\textstyle{\frac{8 \pi \kappa}{c^2}} \mu $,
$-\textstyle{\frac{8 \pi \kappa}{c^2}} \bar \mu$,
$2 \, \phi^{.\gamma} \phi_{.\gamma}$, and
$-2 \, \psi^{.\gamma} \psi_{.\gamma}$
in the action can be seen as allowing `relaxations' of the metric, analogously
to accommodating `constraints' on the metric by including Lagrange multipliers
in the action (e.~g., $\lambda$~in the action
$\bigintsss \! ({\bm R} + \lambda) \, |g|^{\frac12} d^n\!x =
\bigintsss \!\! {\bm R} \, |g|^{\frac12} d^n\!x +
\lambda \bigintsss \!\! |g|^{\frac12} d^n\!x$ for an $n$-dimensional Einstein
manifold, to accommodate the constraint that the volume
$\bigintsss \!\! |g|^{\frac12} d^n\!x$ of the integration region is held fixed).
Just as one does not vary Lagrange multipliers, one should not vary $\mu$,
$\bar \mu$, $\phi$, or $\psi$.  In this role $\mu$, $\bar \mu$,
$\phi$, and $\psi$ simply relax the field equations to allow a larger class of
metrics to satisfy them.  The relaxed field equations will be as useful as the
metrics that satisfy them, no more, no less.

Breaking the scalar field portion of the action into two parts, one
($2 \, \psi^{.\gamma} \psi_{.\gamma}$) coupled to geometry with the orthodox
polarity, the other ($2 \, \phi^{.\gamma} \phi_{.\gamma}$) coupled with the
opposite (`phantom' or `ghost') polarity is justified by the absence of any
compelling reason for choosing one coupling over the other.  The usual mantra
accompanying the making of such a choice is that a scalar field coupled to
geometry with the `wrong sign' has `negative energy' and therefore represents
`exotic matter' that can exist if at all only in highly contrived
circumstances.  This misconception traces back to Einstein's mistaking as
active the passive-inertial density $\rho$ in his ``energy-tensor of matter
${\text T}^\alpha_\sigma$\,''.  If his field equations
{${\bm R}_{\alpha \beta} - \frac12 {\bm R} g_{\alpha \beta}
= \frac{8 \pi \kappa}{c^2} T_{\alpha \beta}$} were changed to
{${\bm R}_{\alpha \beta} - \frac12 {\bm R} g_{\alpha \beta}
= -\frac{8 \pi \kappa}{c^2} T_{\alpha \beta}$}, the effect, in the absence of
pressure, would be equivalent to taking the density $\rho$ to be negative in
the original equations, in which case the matter it purports to represent would
be gravitationally repulsive, and `exotic' for having negative inertial mass
and therefore negative energy.  Correcting Einstein's mistake reveals the
constructions `exotic matter with negative energy density', `phantom field',
`ghost field', and others like them to be little more than instances of
misleading jargon.

Variation of the metric in the action integral of Eq.~(\ref{eqn6}) generates
the field equations
\begin{equation}
{\bm R}_{\alpha \beta} - \textstyle{\frac12} {\bm R} \, g_{\alpha \beta}
  = \frac{8 \pi \kappa}{c^2} T_{\alpha \beta}
\label{eqn7}
\end{equation}
with
\vspace{1\baselineskip}
\begin{equation}
\begin{split}
\frac{8 \pi \kappa}{c^2} T_{\alpha \beta} \defeq
           -\frac{4 \pi \kappa}{c^2} (\mu + \bar \mu) \, g_{\alpha \beta}
          &- 2 \, (\phi_{.\alpha} \phi_{.\beta}
           - \textstyle{\frac12} \phi^{.\gamma} \phi_{.\gamma}
             \, g_{\alpha \beta}) \\
          &+ 2 \, (\psi_{.\alpha} \psi_{.\beta}
           - \textstyle{\frac12} \psi^{.\gamma} \psi_{.\gamma}
             \, g_{\alpha \beta}) \, ,
\end{split}
\label{eqn8}
\end{equation}
and, in consequence,
\begin{equation}
2 \, (\square \phi) \phi_{.\alpha} - 2 \, (\square \psi) \psi_{.\alpha}
  \defeq 2 \, \phi^{.\gamma}\!{}_{:\gamma} \phi_{.\alpha}
         - 2 \, \psi^{.\gamma}\!{}_{:\gamma} \psi_{.\alpha}
  = -\textstyle{\frac{4 \pi \kappa}{c^2}} (\mu + \bar \mu)_{.\alpha} \, .
\label{eqn9}
\end{equation}
The latter of these, which follows from the vanishing of the divergence of
$T_{\alpha \beta}$ implied by the canonical (Bianchi identity) vanishing of the
divergence of
${\bm R}_{\alpha \beta} - \textstyle{\frac12} {\bm R} \, g_{\alpha \beta}$
in the former, is what one has in place of the wave equations
$\square \phi = 0$ and $\square \psi = 0$ that would have resulted from varying
$\phi$ and $\psi$.  It leaves $\phi$ and $\psi$ underdetermined, which is
consistent with their roles as `relaxants' to allow a wider range of metrics to
satisfy the field equations than would be allowed in their absence.

The next two sections will test the usefulness of these field equations by
examining a homogeneous cosmological model they admit as a solution.

\section{Cosmic Evolution Equations}
\label{sec4}

For a Robertson--Walker metric $c^2 dt^2 - R^2(t) ds^2$ (with $t$ in seconds,
$s$ dimensionless, $c$ in meters per second, and $R$ in meters) and
dimensionless scalar fields $\phi = \alpha(t)$ and $\psi = \beta(t)$ the field
Eqs.~(\ref{eqn7}) reduce to
\begin{align}
3 \, \frac{\dot R^2/c^2 + k}{R^2}
 &= -\frac{4 \pi \kappa}{c^2} (\mu + \bar \mu)
     - \frac{\dot \alpha^2 - \dot \beta^2}{c^2} \, ,
\label{eqn10}
\intertext{\vskip -5pt \noindent and \vskip -5pt}
\frac{2}{c^2} \frac{\ddot R}{R} + \frac{\dot R^2/c^2 + k}{R^2}
 &= -\frac{4 \pi \kappa}{c^2} (\mu + \bar \mu)
     + \frac{\dot \alpha^2 - \dot \beta^2}{c^2} \, ,
\label{eqn11}
\end{align}
\noindent
where $k$ = $-1$, 0, or 1, the uniform curvature of the spatial metric $ds^2$.
These equations, which are replacements for the well-studied
Friedmann--Lema\^itre cosmological
equations, are equivalent together to
\begin{align}
\frac{1}{c^2} \frac{\dot R^2}{R^2}
 &= -\frac{4 \pi \kappa}{3 c^2} (\mu + \bar \mu) - \frac{k}{R^2}
    - \frac{\dot \alpha^2 - \dot \beta^2}{3 c^2}
\label{eqn12} \\
\intertext{\vskip -5pt \noindent and \vskip -5pt}
\frac{1}{c^2} \frac{\ddot R}{R}
 &= -\frac{4 \pi \kappa}{3 c^2} (\mu + \bar \mu)
    + \frac{2 (\dot \alpha^2 - \dot \beta^2)}{3 c^2} \, .
\label{eqn13}
\end{align}
\noindent
Equations~(\ref{eqn9}) reduce to
\begin{equation}
\frac{2}{c^2} \left(\ddot \alpha
                    + 3 \, \frac{\dot R}{R} \dot \alpha \right) \dot \alpha -
\frac{2}{c^2} \left(\ddot \beta
                    + 3 \, \frac{\dot R}{R} \dot \beta \right) \dot \beta
  = -\frac{4 \pi \kappa}{c^2} \partial_t (\mu + \bar \mu)
\label{eqn14}
\end{equation}
\noindent
for the time component and $\partial_{a}(\mu + \bar \mu) = 0$ for the
space components, thus impose spatial but not temporal uniformity on
$\mu + \bar \mu$.  Equation~(\ref{eqn14}) is equivalent to
\begin{equation}
\frac{1}{c^2} \left(R^6 (\dot \alpha^2 - \dot \beta^2)\right)^{\bm .}
  = -\frac{4 \pi \kappa}{c^2} R^6 \partial_t (\mu + \bar \mu) \, .
\label{eqn15}
\end{equation}

To give substance to the densities $\mu$ and $\bar \mu$ let us invoke, as
assumed ingredients of the model under construction, two kinds of gravitating
matter: a) {\it primordial matter}, existent at all times, never changing in
amount, with its net density $\mu + \bar \mu$ thus inversely proportional to
the cube of the scale factor $R$; and b) {\it continuously created} (or
{\it destroyed}) {\it matter}, coming steadily into existence (or passing out
of existence) at a rate just sufficient to keep its net density constant.  The
first of these is, except for the inclusion of a repulsive component of the
density, not essentially different from what one assumes for the $\Lambda$CDM
model.  The second is essentially the same as the basic assumption on which the
so-called `steady state' cosmology is built, also excepting the repulsive
density component.  Incorporating both will allow our model to provide a good
fit to relevant astronomical observations. 

For the net density of primordial matter we have
\begin{equation}
\mu_{\text P}(t) + \bar \mu_{\text P}(t)
  = (\mu_{{\text P},0} + \bar \mu_{{\text P},0}) \frac{R^3(t_0)}{R^3(t)} \, ,
\label{eqn16}
\end{equation}
where $t_0$ denotes the value of $t$ at the present epoch, and
$\mu_{{\text P},0}$ and $\bar \mu_{{\text P},0}$ are the present values of the
primordial densities $\mu_{\text P}$ and $\bar \mu_{\text P}$.  For the net
density of continuously created matter we have
\begin{equation}
\mu_{\text C}(t) + \bar \mu_{\text C}(t)
  = \mu_{{\text C},0} + \bar \mu_{{\text C},0} \, ,
\label{eqn17}
\end{equation}
where $\mu_{{\text C},0}$ and $\bar \mu_{{\text C},0}$ are the present values
of the continuously created densities $\mu_{\text C}$ and $\bar \mu_{\text C}$.
Under the assumption that the net densities
$\mu_{\text P} + \bar \mu_{\text P}$ and
$\mu_{\text C} + \bar \mu_{\text C}$ are additive, i. e.,
$\mu + \bar \mu = (\mu_{\text P} + \bar \mu_{\text P}) +
                       (\mu_{\text C} + \bar \mu_{\text C})$,
Eq.~(\ref{eqn15}) turns into
\begin{equation}
\frac{1}{c^2} \!\! \left(R^6 (\dot \alpha^2 - \dot \beta^2)\right)^{\bm .}
  = \frac{4 \pi \kappa}{c^2}
    (\mu_{{\text P},0} + \bar \mu_{{\text P},0})
    \, R^3(t_0) \left(R^3\right)^{\bm {. {}}} \, ,
\label{eqn18}
\end{equation}
which integrates to
\begin{equation}
\frac{1}{c^2} (\dot \alpha^2 - \dot \beta^2)
  = \frac{4 \pi \kappa}{c^2}
    (\mu_{{\text P},0} + \bar \mu_{{\text P},0}) \frac{R^3(t_0)}{R^3}
    + \frac{B}{R^6} \, ,
\label{eqn19}
\end{equation}
where $B$ is the constant of integration, with units m$^4$.

At this point it is convenient to set
\begin{align}
A_{\text P}
 &\defeq - \frac{4 \pi \kappa}{c^2}
           (\mu_{{\text P},0} + \bar \mu_{{\text P},0}) \, R^3(t_0)
  \qquad\text{(units m)}
\label{eqn20}
\intertext{and}
A_{\text C}
 &\defeq  - \frac{4 \pi \kappa}{c^2}
            (\mu_{{\text C},0} + \bar \mu_{{\text C},0})
  \qquad\text{(units m$^{-2}$)}.
\label{eqn21}
\end{align}
Equation~(\ref{eqn19}) then becomes
\begin{equation}
\frac{1}{c^2} (\dot \alpha^2 - \dot \beta^2)
  = \frac{B - A_{\text P} R^3}{R^6} \, ,
\label{eqn22}
\end{equation}
and substitution from Eqs.~(\ref{eqn20}), (\ref{eqn21}), and (\ref{eqn22}) into
Eqs.~(\ref{eqn12}) and~(\ref{eqn13}) produces
\vskip 0.25\baselineskip
\begin{align}
\frac{1}{c^2} \frac{\dot R^2}{R^2}
 &= \frac{P_1(R)}{3 R^6}
  = \frac{A_{\text C}}{3} - \frac{k}{R^2}
    + \frac{2 A_{\text P}}{3 R^3} - \frac{B}{3 R^6}
\label{eqn23} \\
\intertext{\vskip -5pt \noindent and \vskip -5pt}
\frac{1}{c^2} \frac{\ddot R}{R}
 &= \frac{P_2(R)}{3 R^6}
  = \frac{A_{\text C}}{3}
    - \frac{A_{\text P}}{3 R^3} + \frac{2 B}{3 R^6} \, ,
\label{eqn24}
\end{align}
where
\begin{align}
P_1(R) &\defeq A_{\text C} R^6 - 3 k R^4 + 2 A_{\text P} R^3 - B
\vspace{-10pt}
\label{eqn25}
\intertext{\noindent and}
P_2(R) &\defeq A_{\text C} R^6 - A_{\text P} R^3 + 2 B \, .
\label{eqn26}
\end{align}

According to Eqs.~(\ref{eqn20}) and (\ref{eqn21}), positive values for
$A_{\text P}$ and $A_{\text C}$ correspond to negative net densities for
primordial matter and continuously created matter, signifying that on balance
matter of each kind gravitationally repels all other matter more strongly than
it attracts other matter.  This excess of repulsion over attraction will be
presumed to exist for both types.  With thus $A_{\text P} > 0$ and
$A_{\text C} > 0$ assumed, along with an added assumption that $B > 0$, several
properties of the scale factor $R$ as a solution of these equations can be
inferred rather easily, to wit:

\begin{itemize}

\item
For each of $k = -1$, 0, 1 there is a least positive number $R_{\text{min}}$
(the least positive root of the polynomial $P_1(R)$) that $R$ cannot go below
without violating Eq.~(\ref{eqn23}), and at which $\dot R = 0$.
(See Fig.~\ref{fig1}.) This sets a positive lower bound on the compression of
space and thereby rules out the development of a `big bang' singularity.

\item
For $k = 1$ and some choices of $A_{\text C}$, $A_{\text P}$, and $B$
(Fig.~\ref{fig1}, graphs~b, c, and d) there are in addition to $R_{\text{min}}$
other values $R_{\text{max}}$ and $R_{\text{min*}}$ of $R$ at which $P_1$, and
consequently $\dot R$, vanish.

\item The identity $P_2(R) = \frac12 R P_1'(R) - 2 P_1(R)$ has the consequence
that because, with one exception (Fig.~\ref{fig1}, graph~b),
$P_1'(R_{\text{min}}) > 0$, also $P_2(R_{\text{min}}) > 0$, and therefore
$\ddot R$ is positive when $R = R_{\text{min}}$.  This tells that in place of a
`big bang' there is a `bounce' off a state of maximum compression at the time
when $R = R_{\text{min}}$.  The lone exception has
$k = 1$, $P_2(R_{\text{min}}) = P_1'(R_{\text{min}}) = 0$, and
$R(t) = R_{\text{min}} = R_{\text{max}}$ for all time, modeling a static,
spherical universe of radius $R_{\text{min}}$.  A similar model
(Fig.~\ref{fig1}, graph~d) has $R(t) = R_{\text{min*}} = R_{\text{max}}$ for
all time, and also as an asymptotic limit from below and from above.

\item With the time $t = 0$ chosen so that $R(0) = R_{\text{min}}$, $R(t)$ is
symmetric about $t = 0$, the evolution of the universe after time zero thus
being mirrored in reverse as a devolution of the universe before time zero.

\item
Whereas for $k = -1$ and $k = 0$ (Fig.~\ref{fig1}), and $k = 1$
(Fig.~\ref{fig1}, graph~a), the universe expands from $R_{\text{min}}$ to
infinity as $t \to \pm \infty$, other behaviors are possible when $k = 1$, most
notably: (Fig.~\ref{fig1}, graph~b) static spherical universe as noted above;
(Fig.~\ref{fig1}, graph~c) spherical universe oscillating for all time between
minimum radius $R_{\text{min}}$ and maximum radius $R_{\text{max}}$.

\begin{figure}
\includegraphics[width=5.0in,height=2.25in]{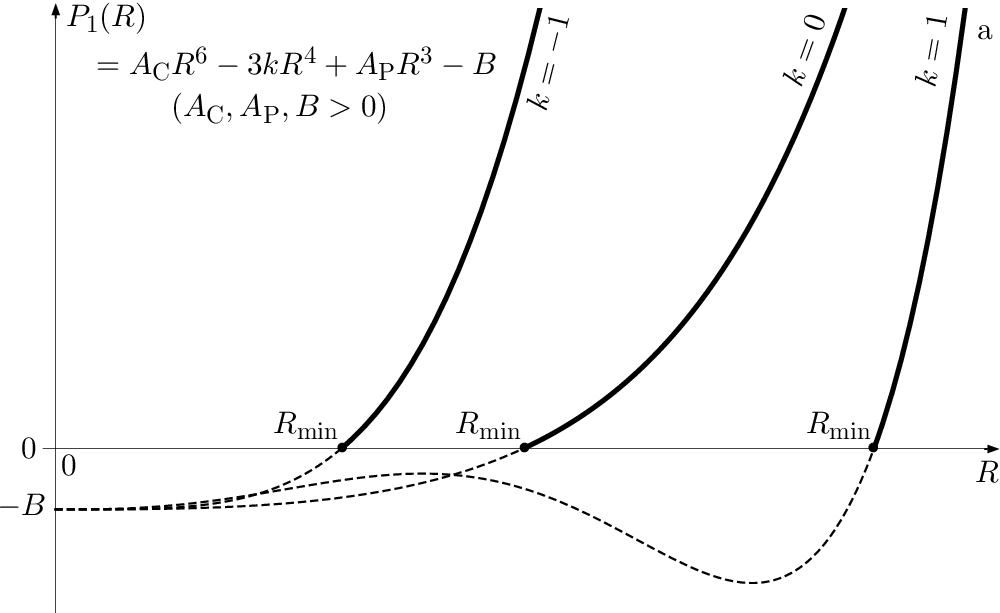}
\vskip 10pt
\includegraphics[width=5.0in,height=2.25in]{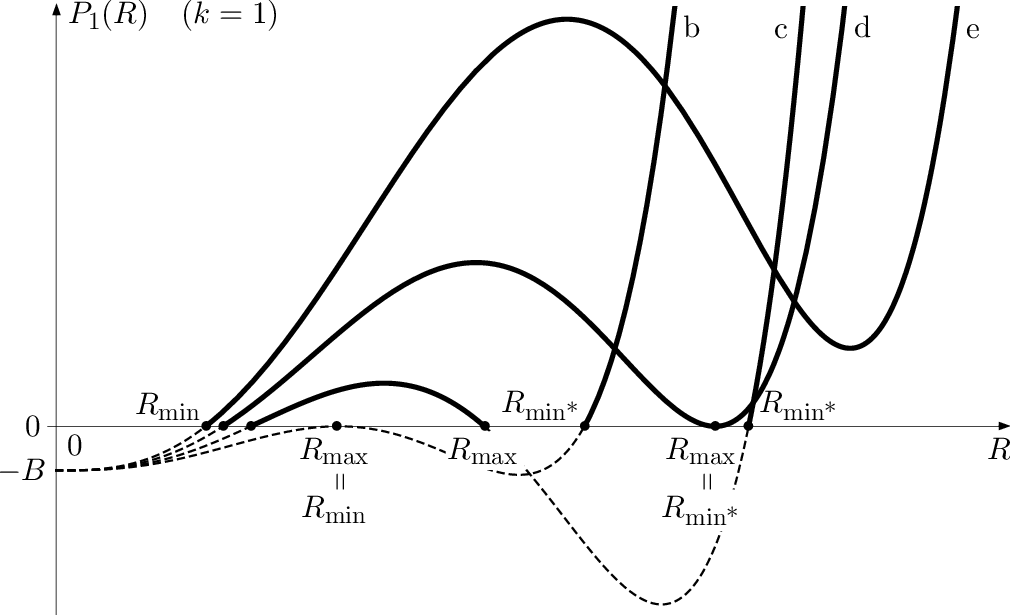}
\caption{\label{fig1} Graphs of $P_1(R)$ for $k = -1$, 0, and 1, and generic
positive values of the parameters $A_{\text C}, A_{\text P}$, and $B$.  Values
of $R$ for which $P_1(R) < 0$ are excluded from the range of $R(t)$ by
Eq.~(\ref{eqn23}).}
\end{figure}

\item The polynomial $P_2(R)$ is quadratic in $R^3$, with discriminant
$A_{\text P}^2 - 8 A_{\text C} B$.  If $B > A_{\text P}^2/8 A_{\text C}$, then
$P_2(R)$ has no positive root, so $\ddot R > 0$ at all times, thus the
universal expansion described by $R(t)$ for times after $t = 0$ is always
accelerating.  If $B = A_{\text P}^2/8 A_{\text C}$, then
$P_2(R) = A_{\text C} (R^3 - A_{\text P}/2 A_{\text C})^2$, so the expansion
is, except for a momentary pause when $R(t)$ passes through
$\sqrt[3]{A_{\text P}/2 A_{\text C}}$, accelerating at all positive times,
unless, as in Fig.~\ref{fig1}, graph~b,
$R(t) = R_{\text{min}} = R_{\text{max}}$ or, as in Fig.~\ref{fig1}, graph~d,
$R(t) = R_{\text{min*}} = R_{\text{max}}$.  If
$B < A_{\text P}^2/8 A_{\text C}$, then $P_2(R)$ has two positive roots,
\begin{equation}
R_{\text d}
 \defeq \left(\frac{A_{\text P}}{2 A_{\text C}}
              - \frac{\sqrt{A_{\text P}^2
                      - 8 A_{\text C} B}}
                     {2 A_{\text C}}\right)^\frac{1}{3},
\label{eqn27}
\end{equation}
which marks a transition from the initial accelerating expansion associated
with the bounce at $t = 0$ to an interval of decelerating expansion, and
\begin{equation}
R_{\text a}
 \defeq \left(\frac{A_{\text P}}{2 A_{\text C}}
              + \frac{\sqrt{A_{\text P}^2
                      - 8 A_{\text C} B}}
                     {2 A_{\text C}}\right)^\frac{1}{3},
\label{eqn28}
\end{equation}
which marks a return to accelerating expansion (if the expansion is destined
to continue forever, as opposed to some of the other possible behaviors when
$k = 1$).

\item For the infinitely expanding models the `Hubble parameter'
$H$ ($ \! \defeq \dot R/R$) and the `acceleration parameter'
$Q$ ($ \! \defeq (\ddot R/R)/(\dot R/R)^2$)
behave asymptotically as follows:
\begin{align}
\frac{1}{c^2} H^2 = \frac{P1(R)}{3 R^6}
                 &= \frac{A_{\text C}}{3} - \frac{3 k R^4
                          - 2 A_{\text P} R^3 + B}{3 R^6}
\label{eqn29} \\
                  &\to \frac{A_{\text C}}{3} \quad
                   \begin{cases} \text{from below if } k > 0 \\
                                 \text{from above if } k \leq 0
                   \end{cases}
                    \hskip -14pt \Bigg\} \; \text{ as } R \to \infty
\label{eqn30}
\intertext{(telling that, for some number $C$,
$R(t) \sim C e^{\pm \sqrt{A_{\text C}/3} \, c \, t}$ as $t \to \pm \infty$) and}
Q = \frac{P_2(R)}{P_1(R)}
                 &= c^2 \, \frac{A_{\text C} R^6 - A_{\text P} R^3
                                 + 2 B}{3 H^2 R^6}
\label{eqn31} \\[5pt]
                 &= 1 + c^2 \, \frac{k R^4 - A_{\text P} R^3 + B}{H^2 R^6}
\label{eqn32} \\
                 &\to 1 \quad
                  \begin{cases} \text{from above if } k > 0 \\
                                \text{from below if } k \leq 0
                  \end{cases}
                   \hskip -14pt \Bigg\} \; \text{ as } R \to \infty \, .
\label{eqn33}
\end{align}

\item For $k = -1$ or 0, and for some cases of $k = 1$, $H$ has a maximum value
$H_{\text{max}}$ at $R = R_{H_{\text{max}}}$, where
$dH/dR = c^2 (k R^4 - A_{\text P} R^3 + B)/H R^7 =~0$.  If $k = 0$,
$R_{H_{\text{max}}} = \sqrt[3]{B/A_{\text P}}$ and
$H_{\text{max}} \! = c \, \sqrt{A_{\text C}/3 + A_{\text P}^2/3 B}$.  As seen
in Fig.~\ref{fig2}, $H$ rises sharply from~0 at $R_{\text{min}}$ to
$H_{\text{max}}$ at $R_{H_{\text{max}}}$, then reverses and tails off
asymptotically to $c \, \sqrt{A_{\text C}/3}$.  One can show that
$R_{\text{min}} \sim \sqrt[3]{B/2 A_{\text P}}$,
$R_{H_{\text{max}}} \sim \sqrt[3]{B/A_{\text P}}$,
$R_{\text d} \sim \sqrt[3]{2 B/A_{\text P}}$, and
$H_{\text{max}} \sim c \, A_{\text P}/\sqrt{3 B}$, as $B \to 0$ with
$A_{\text C}$ and $A_{\text P}$ fixed.  Thus, as $B \to 0$ with $A_{\text C}$
and $A_{\text P}$ fixed, $R_{\text{min}}$, $R_{H_{\text{max}}}$, and
$R_{\text d}$ are squeezed together closer and closer to 0, and
$H_{\text{max}}$ grows without bound.  This clearly is a recipe for an
explosive postbounce inflation ending with a `graceful exit' initiated by the
onset of deceleration when $R$ reaches~$R_{\text d}$.

\begin{figure}[b]
\includegraphics[width=5.0in,height=2.25in]{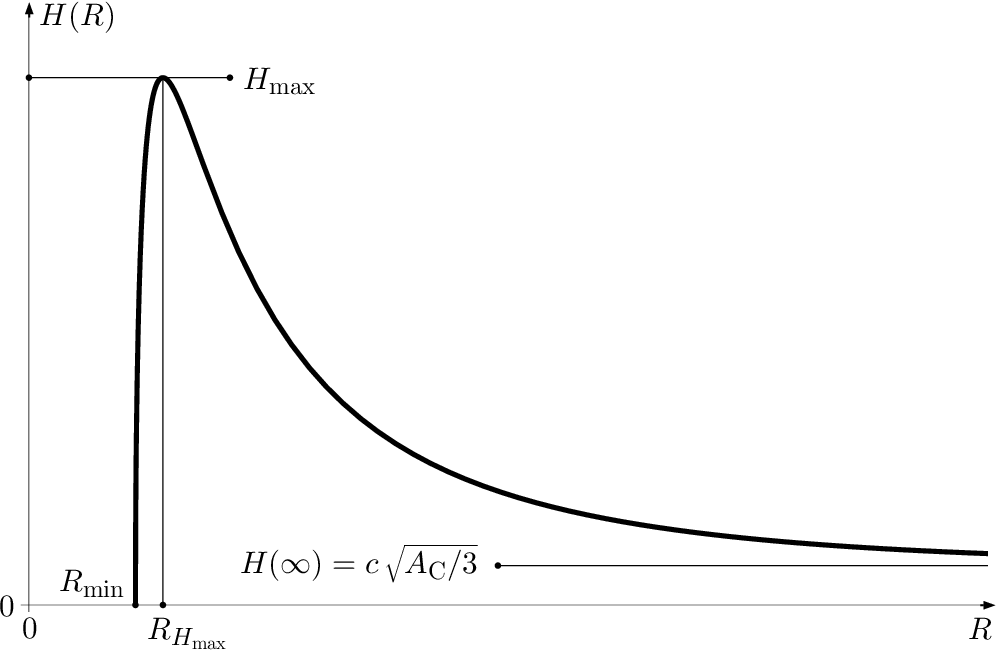}
\vskip 20pt
\includegraphics[width=5.0in,height=2.25in]{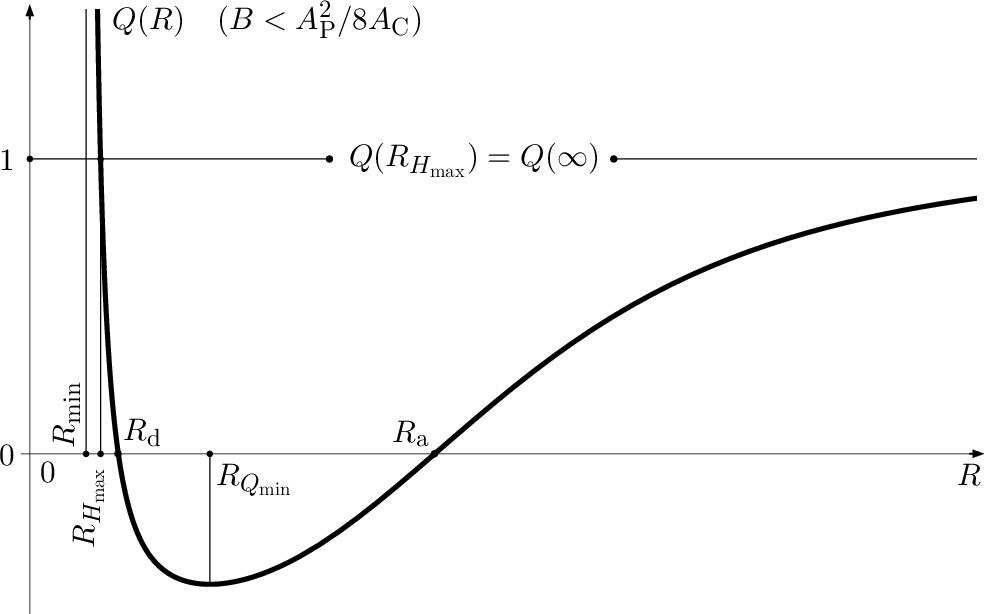}
\caption{\label{fig2} Graphs of $H(R)$ and $Q(R)$ for $k = -1$ or 0 and generic
positive values of the parameters $A_{\text C}$, $A_{\text P}$, and $B$, and
for some cases of $k = 1$, showing early stage inflation followed by a decline
in $H$ to its asymptotic limit $H(\infty) = c \sqrt{A_{\text C}/3}$, and (for
$B < A_{\text P}^2/8 A_{\text C}$) transitions of $Q$ from inflationary
acceleration to deceleration at $R_{\text d}$ and back to acceleration at
$R_{\text a}$ to the asymptotic limit $Q(\infty) = 1$.}
\end{figure}

\begin{figure}[b]
\includegraphics[width=5.0in,height=2.25in]{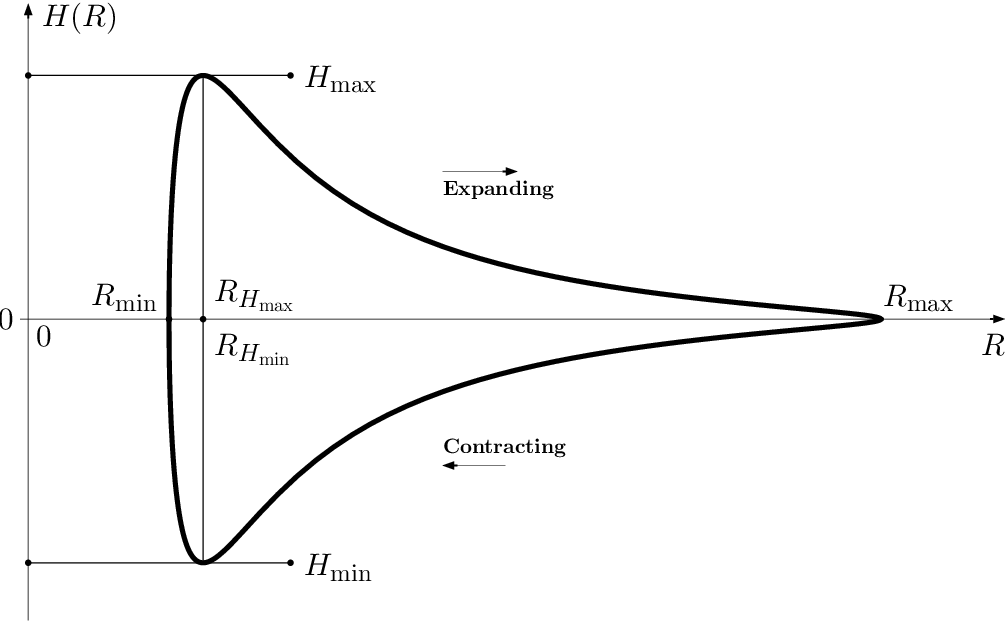}
\vskip 20pt
\includegraphics[width=5.0in,height=2.25in]{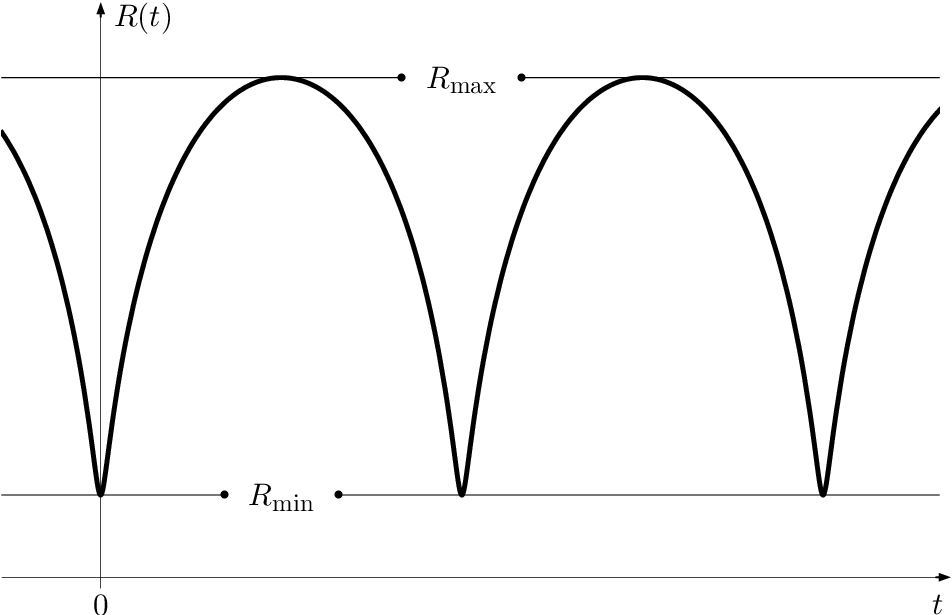}
\caption{\label{fig3} Graphs of $H(R)$ and $R(t)$ for $k = 1$ and generic
positive values of the parameters $A_{\text C}$, $A_{\text P}$, and $B$
associated with graph c of Fig.~\ref{fig1}, showing repetitive, identical
periods of expansion and contraction, each beginning with a stage of rapid
inflation from a bounce at $R = R_{\text{min}}$, which is followed by a less
rapid expansion to $R = R_{\text{max}}$, then a mirror-image contraction to an
ending stage of rapid deflation into the next bounce at $R = R_{\text{min}}$.
The graphed functions are related by $\dot R(t)/R(t) \eqdef H(R(t))$.}
\end{figure}

\item As Fig.~\ref{fig2} exhibits, for $k = -1$, $k = 0$, and $k = 1$, graph a,
the acceleration parameter $Q$, descending from $\infty$ at $R_{\text{min}}$,
passes through 1 at $R_{H_{\text{max}}}$, bottoms out with a minimum value
$Q_{\text{min}}$ at $R_{Q_{\text{min}}}$, where $dQ/dR = 0$,
then creeps slowly back to 1 as $R \to \infty$ (with a late pass through 1 if
$k = 1$).  When $k = 0$,
$Q_{\text{min}}
= -\frac{1}{2} + \frac{3}{2} \sqrt{A_{\text C} B/(A_{\text P}^2
   + A_{\text C} B})$
at
$R_{Q_{\text{min}}}
= \left(B/A_{\text P} + \sqrt{B^2/A_{\text P}^2
   + B/A_{\text C}}\right)^\frac{1}{3}$, which, with
$A_{\text C}$ and $A_{\text P}$ fixed, are asymptotic respectively to
$-\frac12$ and $\sqrt[6]{B/A_{\text C}}$ as $B \to 0$.

\item If $k = 1$ and $A_{\text C}$, $A_{\text P}$, and $B$ are such as to
generate graph c of Fig.~\ref{fig1}, then for the section bounded by
$R_{\text{min}}$ and $R_{\text{max}}$ the behavior of $R$ is cyclical, as
Fig.~\ref{fig3} portrays graphically.

\end{itemize}

\section{Cosmological Solutions}
\label{sec5}

Study of solutions of the field equations (23) and (24) proceeds most smoothly
when they are recast in terms of the variable $U \defeq S^3$, where
$S \defeq R/R_{\text{min}}$.  The recast equations are together equivalent to
\begin{equation}
\ddot U = 3 c^2 A_{\text C} U - \frac{6 c^2 k}{R_{\text{min}}^2} U^\frac{1}{3}
          + \frac{3 c^2 A_{\text P}}{R_{\text{min}}^3}
\label{eqn34}
\end{equation}
and an equation for $\dot U^2$ that is essentially redundant.  Initial
conditions for $U$ are $U(0) = 1$ and $\dot U(0) = 0$, obtained from
$S(0) = R(0)/R_{\text{min}} = 1$ and
$\dot S(0) = \dot R(0)/R_{\text{min}} = 0$.

\subsection{Flat open universe ($k = 0$)}
\label{sec5.1}

\begin{figure}
\includegraphics[width=5.0in,height=2.25in]{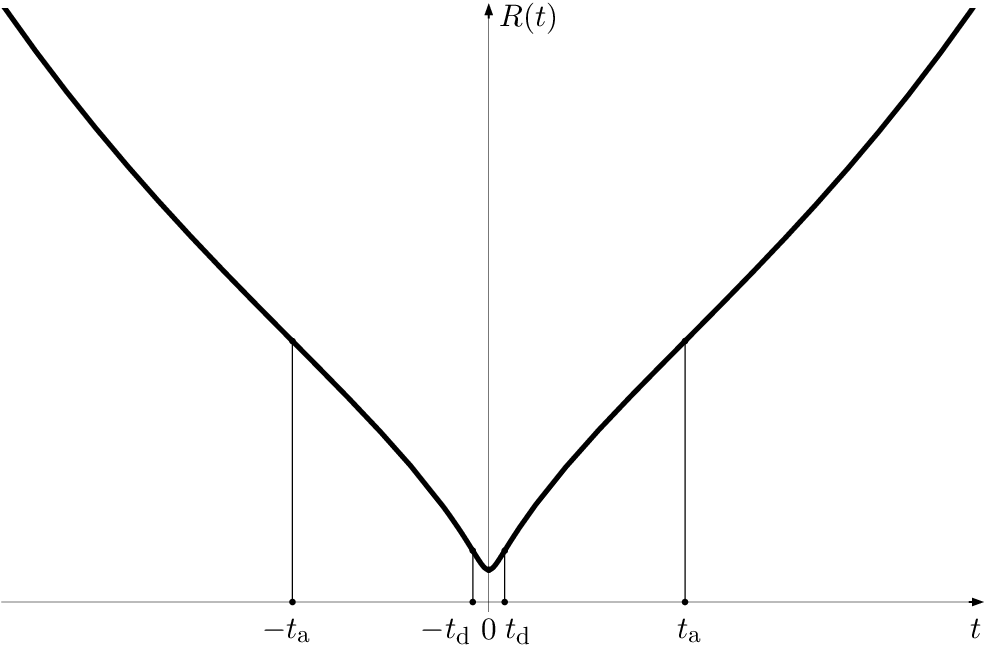}
\vskip 20pt
\includegraphics[width=5.0in,height=2.25in]{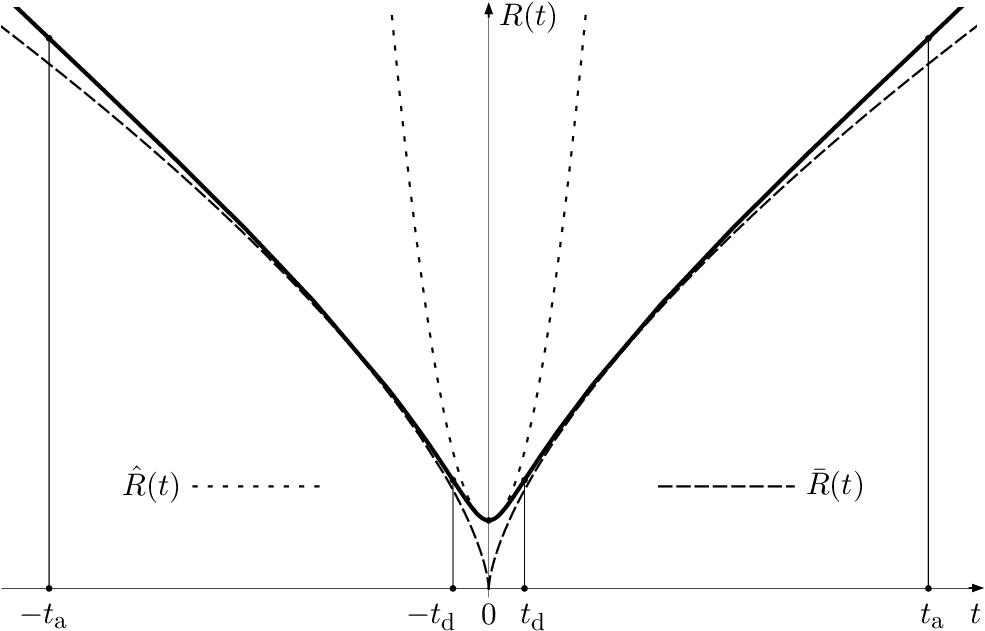}
\vskip 5pt
\caption{\label{fig4} Graph of $R(t)$ for $k = 0$ and generic positive values
of $A_{\text C}$, $A_{\text P}$, and $B$ with
$B < A_{\text P}^2/8 A_{\text C}$, showing early acceleration from $t = 0$ to
$t = t_{\text d}$, deceleration from $t_{\text d}$ to $t_{\text a}$, and
ultimately exponential acceleration after $t_{\text a}$; truncated graph of
$R(t)$ with approximations $\hat R(t)$ and $\bar R(t)$ as identified in
Eqs.~(\ref{eqn38}) and~(\ref{eqn40}).}
\end{figure}

When $k = 0$, so that space is perfectly flat,
$R_{\text{min}}^3 =
-A_{\text P}/A_{\text C} + \sqrt{A_{\text P}^2 + A_{\text C} B}/A_{\text C}$
and it is straightforward to integrate Eq.~(\ref{eqn34}), the result being,
upon reversion from $U$ to $R$,
\begin{equation}
R^3(t) = \left(R_{\text{min}}^3 + \frac{A_{\text P}}{A_{\text C}}\right)
              \cosh \left(\sqrt{3 A_{\text C}} \, c \, t\right)
         - \frac{A_{\text P}}{A_{\text C}} \, .
\label{eqn35}
\end{equation}
One can show that if $B \leq A_{\text P}^2/8 A_{\text C}$, so that
$R_{\text d}$ and $R_{\text a}$ exist, then
$R_{\text{min}} < R_{\text d} \leq R_{\text a}$.  The postbounce times
$t_{\text d}$ and $t_{\text a}$ at which $R = R_{\text d}$ and
$R = R_{\text a}$ are the positive solutions of
\begin{equation}
\cosh \left(\sqrt{3 A_{\text C}} \, c \, t_{\{\text{d,a}\}} \right)
 = \frac{{R_{\{\text{d,a}\}}^3} + A_{\text P}/A_{\text C}}
        {R_{\text{min}}^3 + A_{\text P}/A_{\text C}} \, .
\label{eqn36}
\end{equation}
The first graph in Fig.~\ref{fig4} displays the evolution of $R(t)$ for generic
values of $A_{\text C}$, $A_{\text P}$, and $B$ with
$B < A_{\text P}^2/8 A_{\text C}$, showing the acceleration during the interval
from $t = 0$ to $t = t_{\text d}$, the deceleration during the interval from
$t_{\text d}$ to $t_{\text a}$, and the ultimately exponential acceleration
after $t_{\text a}$.

For times $t \approx 0$, Eq.~(\ref{eqn35}) yields the approximation
\begin{equation}
R^3(t) \approx R_{\text{min}}^3
               \left(1 + \frac{3 A_{\text C}}{2}
                         \frac{R_{\text{min}}^3
                               + A_{\text P}/A_{\text C}}
                              {R_{\text{min}}^3} c^2 t^2 \right) \, ,
\label{eqn37}
\end{equation}
which produces the further approximation that
\begin{equation}
R(t) \approx \hat R(t)
     \defeq R_{\text{min}}
            + \frac{A_{\text C}}{2}
              \left(R_{\text{min}}
                    + \frac{A_{\text P}}{A_{\text C}}
                      \frac{1}{R_{\text{min}}^2} \right) c^2 t^2 \, ,
\label{eqn38}
\end{equation}
thus that $R(t)$ is approximately parabolic, opening upward from a vertex at
$R_{\text{min}}$, with an inflationary `steepness' factor that grows as
$c^2 A_{\text P}/2 R_{\text{min}}^2$ as $R_{\text{min}} \to 0$.

Equivalent to Eq.~(\ref{eqn37}) is
\begin{equation}
R^3(t) - R_{\text{min}}^3
 \approx \frac{3 A_{\text C}}{2}
         \left(R_{\text{min}}^3 + \frac{A_{\text P}}{A_{\text C}} \right)
         (c \, t)^2 \, .
\label{eqn39}
\end{equation}
After $R(t)$ has inflated to the point that $R_{\text{min}}/R(t) \ll 1$,
$(R^3(t) - R_{\text{min}}^3)^\frac{1}{3}$ will differ but little from $R(t)$,
and then, for so long as $t$ remains small enough for the approximate
Eq.~(\ref{eqn37}) to hold, the approximation
\begin{equation}
R(t) \approx \bar R(t)
     \defeq \left[\frac{3 A_{\text C}}{2}
                  \left(R_{\text{min}}^3
                        + \frac{A_{\text P}}{A_{\text C}}
                  \right)\right]^\frac{1}{3} (c \, t)^{2/3}
\label{eqn40}
\end{equation}
will be valid.  The second graph in Fig.~\ref{fig4} shows the approximations
$\hat R(t)$ and $\bar R(t)$.

\subsection{Nonflat open ($k = -1$) and closed ($k = 1$) universes}
\label{sec5.2}

When $k = -1$ or 1, elementary symbolic solutions of Eq.~(\ref{eqn34}) are
unavailable, so one must resort to numerical integration.  To enable the
integration, numerical values must be chosen for the parameters other than $k$
that appear in Eq.~(\ref{eqn34}), namely, $A_{\text C}$, $A_{\text P}$, and $B$
(implicitly through~$R_{\text{min}}$).  An obvious way to effect this
determination is to impose the requirement that the parameters provide a
distance modulus curve that is a good fit to a Hubble diagram based on data
from the observations of type~Ia supernovae that revealed the acceleration of
the expansion of the universe.  Toward that end a number of definitions are in
order:

\begin{itemize}

\item $t_0$, the time of the present epoch, i.~e., the time elapsed since the
bounce at $t = 0$.

\item $H_0 \defeq H(t_0)$, the present value of the Hubble parameter.

\item $R_0 \defeq c/H_0$, usually referred to as the `Hubble radius' and as
the `radius of the visible universe'.

\item $z \defeq R(t_0)/R - 1$, the redshift function.

\item $z_{\text d} \defeq R(t_0)/R_{\text d} - 1$ and
$z_{\text a} \defeq R(t_0)/R_{\text a} - 1$, the redshifts associated with
the transitions from acceleration to deceleration and back to acceleration.

\item $\tilde R \defeq R/R(t_0) = 1/(1 + z)$, a conventional notation.

\item ${\tilde R}_{\text{min}} \defeq R_{\text{min}}/R(t_0)$ and
      ${\tilde R}_{\{\text{d,a}\}} \defeq R_{\{\text{d,a}\}}/R(t_0)$.

\item $\lambda \defeq c/H_0 R(t_0) = R_0/R(t_0)$, a useful parameter.

\item
\begin{align}
\Omega_{\text C} &\defeq \frac{c^2}{3 H_0^2} A_{\text C} \, ,
\label{eqn41} \\
\Omega_{\text k} &\defeq -\frac{c^2}{H_0^2 R^2(t_0)} k
                   = -\lambda^2 k \, ,
\label{eqn42} \\
\Omega_{\text P} &\defeq \frac{2 c^2}{3 H_0^2 R^3(t_0)} A_{\text P}
                   = \frac{2 H_0 \lambda^3}{3 c} A_{\text P} \, ,
\label{eqn43}
\intertext{\vskip -5pt \noindent and \vskip -5pt}
\Omega_{\text B} &\defeq -\frac{c^2}{3 H_0^2 R^6(t_0)} B
                   = -\frac{H_0^4 \lambda^6}{3 c^4} B \, .
\label{eqn44}
\end{align}

\end{itemize}

\noindent
With these definitions Eq.~(\ref{eqn23}) is equivalent to
\begin{equation}
H^2 = H_0^2 \left(\Omega_{\text C} +
                  \Omega_{\text k} \frac{1}{{\tilde R}^2} +
                  \Omega_{\text P} \frac{1}{{\tilde R}^3} +
                  \Omega_{\text B} \frac{1}{{\tilde R}^6}\right) \, ,
\label{eqn45}
\end{equation}
which upon evaluation at $t_0$ yields
\begin{equation}
\Omega_{\text C} + \Omega_{\text k}
                 + \Omega_{\text P}
                 + \Omega_{\text B} = 1 \, .
\label{eqn46}
\end{equation}
On the other hand, Eq.~(\ref{eqn24}) is equivalent to
\begin{align}
\frac{\ddot {\tilde R}}{\! \tilde R}
 &= H_0^2 \left(\Omega_{\text C}
                 - \frac{1}{2} \, \Omega_{\text P} \frac{1}{{\tilde R}^3}
                 - 2 \, \Omega_{\text B} \frac{1}{{\tilde R}^6} \right)
\label{eqn47} \\
 &= H_0^2 \frac{2 \, \Omega_{\text C} {\tilde R}^6
                - \Omega_{\text P} {\tilde R}^3 - 4 \, \Omega_{\text B}}
               {2 \, {\tilde R}^6} \, ,
\label{eqn48}
\end{align}
which yields, as roots of $\ddot {\tilde R} = 0$,
\begin{align}
{\tilde R}_{\text d}
 &\defeq \left(\frac{\Omega_{\text P}}{4 \, \Omega_{\text C}}
               - \frac{\sqrt{\Omega_{\text P}^2
                       + 32 \, \Omega_{\text C} \, \Omega_{\text B}}}
                      {4 \, \Omega_{\text C}}\right)^\frac{1}{3}
\label{eqn49}
\intertext{\vskip -5pt \noindent and \vskip -5pt}
{\tilde R}_{\text a}
 &\defeq \left(\frac{\Omega_{\text P}}{4 \, \Omega_{\text C}}
               + \frac{\sqrt{\Omega_{\text P}^2
                       + 32 \, \Omega_{\text C} \, \Omega_{\text B}}}
                      {4 \, \Omega_{\text C}}\right)^\frac{1}{3}
\label{eqn50}
\end{align}
(obtainable also from Eqs.~(\ref{eqn27}) and~(\ref{eqn28}) by substitution from
Eqs.~(\ref{eqn41}), (\ref{eqn43}), and (\ref{eqn44})).

To produce a curve to match against the data points of a Hubble diagram one
needs the following standard formula for the luminosity distance $D_{\text L}$
of a photon-emitting astronomical object at redshift $z$:
\begin{equation}
D_{\text L}(z)
 = (1 + z) R(t_0) \,
   r_{\text k} \! \left(\frac{c}{R(t_0)}
                        \bigintss_0^z \!\!\!
                           \frac{1}{\overset{\, \ast}{H}(u)} \, du \right) \, .
\label{eqn51}
\vspace{-10pt}
\end{equation}
Here
\begin{align}
\vspace{-10pt}
\hspace{1pt}
r_{\text k}(\rho)
 \defeq \begin{cases}
          \sinh \rho &\text{if} \quad k = -1, \\
          \rho       &\text{if} \quad k = 0, \\
          \sin \rho  &\text{if} \quad k = 1,
        \end{cases}
\hspace{-8pt}
\label{eqn52}
\end{align}
and, from Eq.~(\ref{eqn45}) and the definition of $z$,
\begin{equation}
\overset{\, \ast}{H}(z) \defeq H_0 [\Omega_{\text C} +
                                    \Omega_{\text k} (1 + z)^2 +
                                    \Omega_{\text P} (1 + z)^3 +
                                    \Omega_{\text B} (1 + z)^6]^\frac12 \, .
\label{eqn53}
\end{equation}
\vskip 10pt
\noindent In terms of the parameter $\lambda$ the luminosity distance formula
reads
\begin{equation}
D_{\text L}(z)
 = \frac{(1 + z) c}{\lambda H_0} \,
   r_{\text k} \! \left(\lambda
               \bigintss_0^z \!\!\!
                  \frac{du}{[\Omega_{\text C} +
                             \Omega_{\text k} (1 + u)^2 +
                             \Omega_{\text P} (1 + u)^3 +
                             \Omega_{\text B} (1 + u)^6]^\frac12} \right) \, .
\label{eqn54}
\end{equation}
\vskip 5pt
From $D_L$ one constructs the distance modulus $\bm \mu$, the difference
between the apparent magnitude and the absolute magnitude of the object in
question at redshift~$z$; this reduces in the usual way to
\begin{equation}
{\bm \mu}(z) = 5 \log_{10} \left(\frac{D_L(z)}{10 \, \text{pc}}\right)
             = 25 + \log_{10}\left(\frac{D_L(z)}{1 \, \text{Mpc}}\right) \, .
\label{eqn55}
\end{equation}

From $\tilde R = 1/(1 + z)$ follows $\dot {\tilde R} = -\dot z/(1 + z)^2 =
-(dz/dt)(\tilde R/(1 + z))$, thus $dt/dz = -1/(1 + z)(\dot {\tilde R}/\tilde R) = -1/(1 + z) \overset{\, \ast}{H}(z)$, which upon integration yields
the following formula for the time $T(z)$ elapsed between an event occurring at
redshift $z$ and the present time $t_0$ (when $z = 0$):
\begin{align}
T(z) &= \bigintss_0^z \!\!\! \frac{1}{(1 + u)\overset{\, \ast}{H}(u)} \, du
\label{eqn56} \\
     &= \frac{1}{H_0} \,
        \bigintss_0^z \!\!\!
           \frac{du}{(1 + u)[\Omega_{\text C} +
                             \Omega_{\text k} (1 + u)^2 +
                             \Omega_{\text P} (1 + u)^3 +
                             \Omega_{\text B} (1 + u)^6]^\frac12} \, .
\label{eqn57}
\end{align}

The distance modulus $\bm \mu$ of Eq.~(\ref{eqn55}) is the function to be
fitted to the relative magnitudes data obtained from observations of type~Ia
supernovae.  Using the SNe~Ia data from the 182-member gold sample described in
Sec.~3 of Ref.~\cite{ries1}, the recent estimate $H_0 = 73.8$ km s$^{-1}$
Mpc$^{-1}$ obtained by Riess, et al.~\cite{ries2}, and the Mathematica function
NonLinearModelFit, I have carried out the fitting and reported the outcome in
detail in Ref.~\cite{elli4}.  The best fit, for which
$\chi_{\text{red}}^2 = 0.869$, has
$k = 1$, $\Omega_{\text P} = 0.479$, $\Omega_{\text C} = 0.954$,
$\Omega_{\text k} = -0.433$, $\Omega_{\text B} = -1.20 \times 10^{-196}$,
$\lambda = 0.658$, and $z_{\text a} = 0.585$, $z_{\text d}$ being set somewhat
arbitrarily at $1 \times 10^{65}$ to make the time at which the scale factor
$R$ has doubled one hundred times from $R_{\text{min}}$ come out to
approximately $10^{-35}$~s.  Figure 5 shows the Hubble diagram for this best
fit, as well as for the corresponding fits for $76.2$ and
$71.4$ km s$^{-1}$ Mpc$^{-1}$, the upper and lower limits of the 3.3\%
confidence predicted in~Ref.~\cite{ries2}.  The evolution of the normalized
scale factor $S = R/R_{\text{min}}$ for the best fit is depicted graphically
in~Fig.~\ref{fig6}.

\begin{figure*}[!ht]
\includegraphics[width=4.85in, height=2.6in]{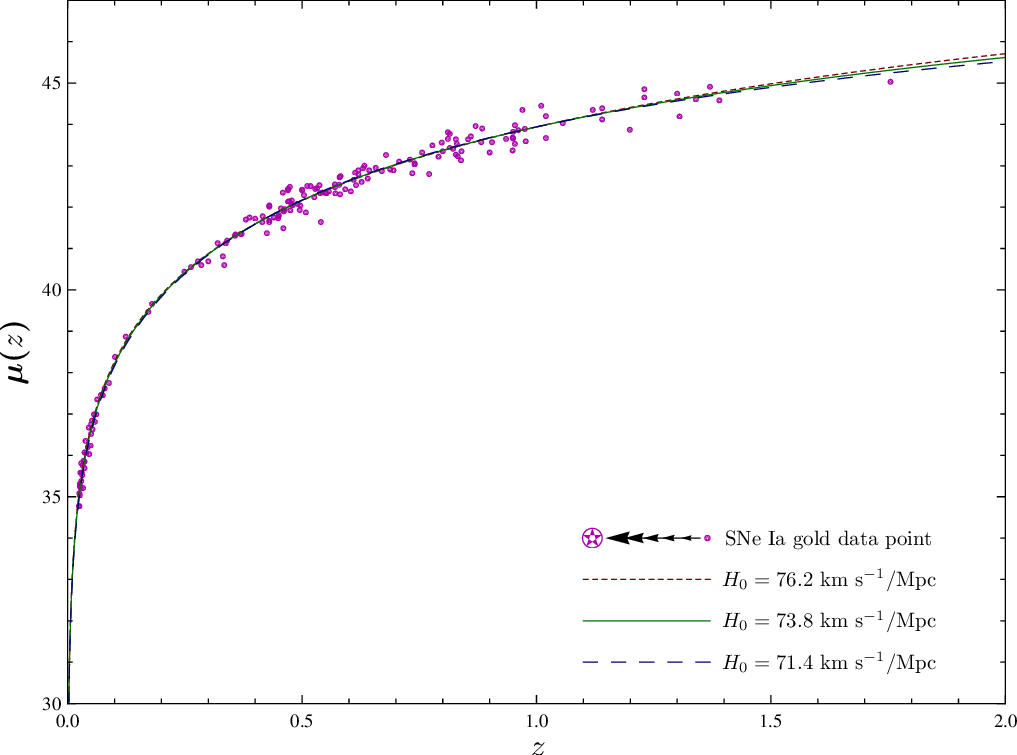}
\caption{\label{fig5} Best-fit models of ${\bm \mu}(z)$ for $k = 1$, plotted
against the 182 SNe~Ia gold sample data points they are fitted to.  The curve
for $H_0 = 73.8$ km s$^{-1}$/Mpc fits the 182 data points with
$\chi_{\text{red}}^2 = 0.869$.}
\end{figure*}

\begin{figure}[!h]
\includegraphics[width=4.85in]{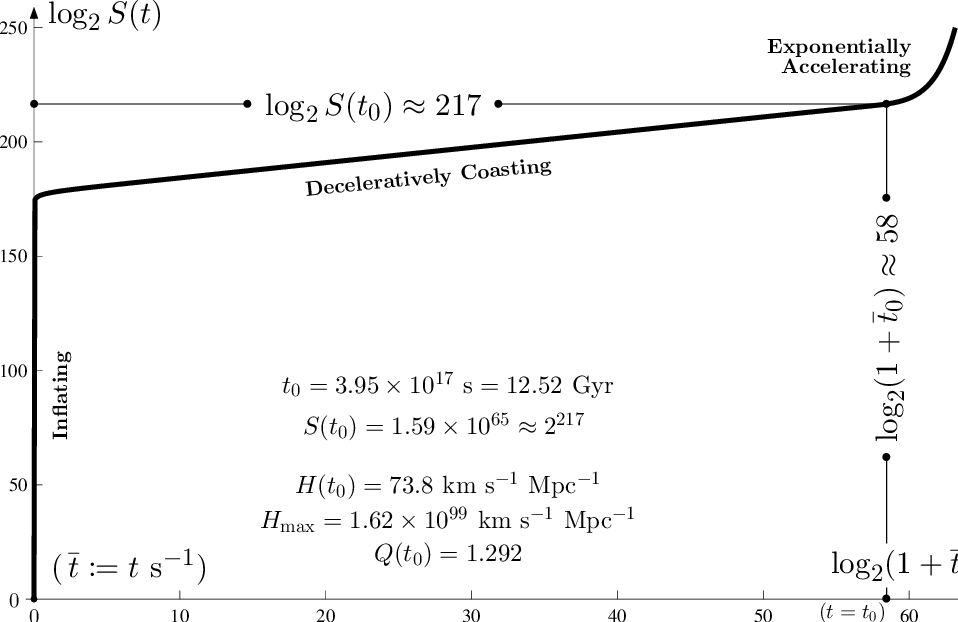}
\caption{\label{fig6} Postbounce graph of $\log_2 S(t)$ versus
$\log_2 (1 + \bar t \,)$ ($\, \bar t \defeq t$ s$^{-1}$) for the
best-fit solution with $H_0 = 73.8$ km s$^{-1}$ Mpc$^{-1}$ and $k = 1$.  The
early stage rapid inflation, after producing 178 doublings of the normalized
scale factor $S$ ($\defeq R/R_{\text{min}}$) in the first second after the
bounce, gives way smoothly to a period of uphill, decelerative `coasting'
(where the graph is nearly linear).  Initially, the acceleration parameter
$Q \defeq (\ddot R/R)/(\dot R/R)^2 = (\ddot S/S)/(\dot S/S)^2$ is positive
and huge ($Q(0) = \infty$).  In much of the coasting period
$Q(t) \approx Q_{\text{min}} \gtrsim -0.5$, which is reflected in the
observation that during that time
$\log_2 S(t) \approx 178 + \frac{217 - 178}{58 - 0} \log_2 (1 + \bar t \,)
             \approx 178 + \frac{2}{3} \log_2 (\bar t\,)$,
so that $S(t) \approx 2^{178} \bar t^{\, \frac{2}{3}}$
($S(t) \propto \bar t^{\,\frac{2}{3}} \implies Q(t) \equiv -\frac{1}{2}$).
After the coasting era $Q(t)$ becomes positive again at redshift
$z_{\text a} = 0.585$ (when $t = 7.06$ Gyr and $\log_2 (1 + \bar t \,) =
57.6$), rising to 1.292 at $t = t_0$, and then to a maximum of 1.310 at
$t = 14.24$ Gyr before settling asymptotically to 1 as $t \to \infty$, the
growth rate of $S(t)$ thus asymptotically becoming exponential as in a de
Sitter universe.  The corresponding graphs for the best-fit solutions with
$H_0$ the same and $k = -1$ and $k = 0$ are not appreciably different from this
one.}
\end{figure}

Perhaps the most surprising prediction of this fitting to SNe~Ia data is that
when Eq.~(\ref{eqn57}) is used to compute the time elapsed since the time of
the `big bounce' the result is 12.5 Gyr.\footnote{For $k = 0$ or $-1$ the time
elapsed is 12.7 Gyr, with  $\chi_{\text{red}}^2 = 0.888$.} This is at variance
with the often quoted interval of 13.7 Gyr for the time elapsed since the
`big bang'.  I would point out, however, that when the corresponding fitting is
carried out for the `standard', or `concordance', $\Lambda$CDM model of
cosmology, with the Friedmann equation
\begin{equation}
H^2 = H_0^2 \left(\Omega_\Lambda +
                  \Omega_{\text k} \frac{1}{{\tilde R}^2} +
                  \Omega_{\text M} \frac{1}{{\tilde R}^3} +
                  \Omega_{\text R} \frac{1}{{\tilde R}^4}\right)
\label{eqn58}
\end{equation}
($M$ for matter, $R$ for radiation) in place of Eq.~(\ref{eqn45}), the best
fit, with the same value of $\chi_{\text{red}}^2$, occurs when $k = 1$,
$\Omega_\Lambda = \Omega_{\text C}$,
$\Omega_{\text k} = \Omega_{\text k}$,
$\Omega_{\text M} = \Omega_{\text P}$, and
$\Omega_{\text R} = 6 \times 10^{-16}$, and also
predicts 12.5 Gyr for the time elapsed since the big bang.  One is left with
the inference that the SNe~Ia data, {\it used alone}, do not support, in either
my model or the $\Lambda$CDM model, the older time of 13.7 Gyr as the `age of
the universe', instead support an age younger by more than a billion years.
This would seem to be a discrepancy that needs addressing.\footnote{An age of
13.2 Gyr can be obtained, with not as good a fit, by taking
$H_0 = 76.2$ km s$^{-1}$ Mpc$^{-1}$, the upper limit of the 3.3\% confidence
interval for $H_0$ predicted in Ref.~\cite{ries2}, and $k = 0$.  At the lower
limit $H_0 = 71.4$ km s$^{-1}$ Mpc$^{-1}$ and the oldest age obtainable is
12.3 Gyr, for $k = 0$ or $-1$.}

\section{Dark Matter and Dark `Energy'}
\label{sec6}

The cosmological model derived in the preceding sections was based on the
premise that the density of the active gravitational mass of all the matter in
the universe is on balance negative.  In the derivation no mechanism for
production of this negative density was specified.  In this section I show how
`drainholes' could provide such a mechanism, and in doing so would explain
dark matter and explain away dark `energy'.

\subsection{Drainholes}
\label{sec6.1}

\begin{figure}
\includegraphics[width=5.0in,height=2.25in]{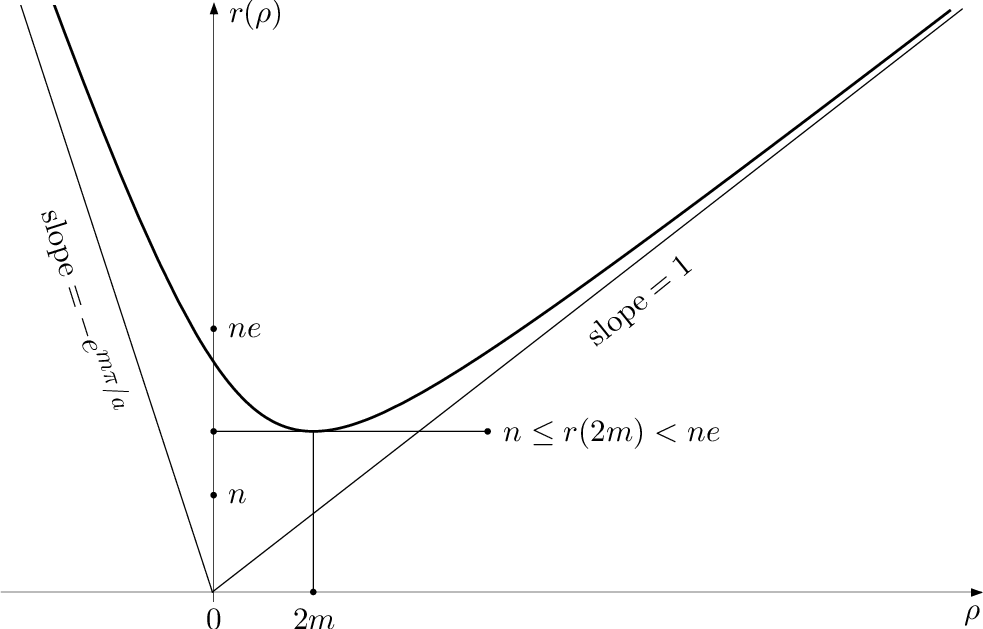}
\vskip 15pt
\includegraphics[width=5.0in,height=2.25in]{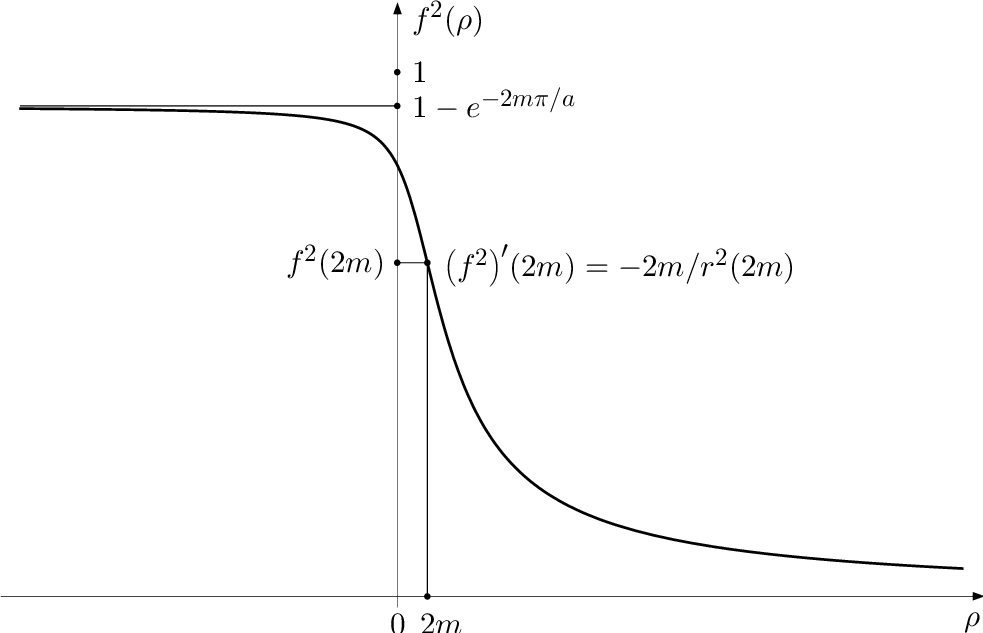}
\caption{\label{fig7} Graphs of $r(\rho)$ and $f^2(\rho)$ for generic values of
the parameters $m$ and $n$ $\left(0 \leq m < n \text{ and }
a \defeq \sqrt{n^2 - m^2}\right)$.}
\end{figure}

The `drainhole' model of a gravitating particle developed in
Ref.~\cite{elli1} and proposed as a source for the negative, gravitationally
repulsive mass density $\bar \mu$ in the action integral of Eq.~(\ref{eqn6}) is
a static, spherically symmetric, vacuum ($\mu = \bar \mu = 0$) solution of the
same field Eqs.~(\ref{eqn7}) that the cosmological model of the preceding
sections is a solution of, but with the unorthodox choice
$\phi \neq 0,\,\psi = 0$.\footnote{${\bm R}_{\alpha \beta}$ and ${\bm R}$ here
are the negatives of those in Ref.~\cite{elli1}.}$^,$\footnote{The scalar
field $\phi$ was presumed in Ref.~\cite{elli1} to satisfy the wave equation
$\square \phi = 0$ obtained from varying $\phi$ in the action integral.
In retrospect that is seen to have been a redundancy, as the field
Eqs.~(\ref{eqn9}), which follow from variation of the metric alone, reduce
when $\psi = 0$ to $2 \, (\square \phi) \phi_{.\alpha} = 0$, a consequence of
which is $\square \phi = 0$.}$^,$\footnote{Apparently the first to find static,
spherically symmetric, vacuum solutions of Eqs.~(\ref{eqn7}), albeit with the
drainhole-excluding, orthodox choice $\phi = 0,\,\psi \neq 0$ (descended
directly from Einstein's assumption that inertial mass produces gravity), was
I.~Z.~Fisher, in 1948.~\cite{fish}}  This space-time manifold has come to be
recognized as an early (apparently the earliest) example of what is now called
by some a `traversable wormhole'~\cite{clem1}, and has been analyzed from
various perspectives by
others~\cite{chcl,clem2,kasabh,arme,shha,perl,daka,nazhza,mull,
dese,dknn,manu,naniiz,abe,naas}.  Its metric has the proper-time forms
\begin{align}
c^2 d\tau^2 &= c^2 dt^2 - [d\rho - f(\rho) \, c \, dt]^2
                        - r^2 (\rho) \, d\Omega^2
\label{eqn59} \\
            &= [1 - f^2(\rho)] \, c^2 dT^2 - \frac{1}{1 - f^2(\rho)} \, d\rho^2
                                           - r^2(\rho) \, d\Omega^2 \, ,
\label{eqn60}
\end{align}
where $T = t + {\displaystyle \frac{1}{c} \! \bigintsss \!\!\!
                              \frac{f(\rho)}{1 - f^2(\rho)} \, d\rho}$,
and with $a \defeq \sqrt{n^2 - m^2}$,
\begin{equation}
f^2(\rho) = 1 - e^{-(2 \, m/n) \phi}
\label{eqn61}
\end{equation}
\noindent and
\vspace{-0.25\baselineskip}
\begin{equation}
  r(\rho) = \sqrt{(\rho - m)^2 + a^2} \, e^{(m/n) \phi}
          = \sqrt{\frac{(\rho - m)^2 + a^2}{1 - f^2(\rho)}} \, ,
\vspace{0.5\baselineskip}
\label{eqn62} \\
\end{equation}
in which
\begin{equation}
      \phi = \alpha(\rho)
           = \frac{n}{a}
             \left[\frac{\pi}{2}
                   - \tan^{-1} \left(\frac{\rho - m}{a}\right)\right] \, ,
\label{eqn63}
\vspace{0.5\baselineskip}
\end{equation}
\noindent
the parameters $m$ and $n$ satisfying $0 \leq m < n$.  (The coordinate $\rho$
used here translates to $\rho + m$ in Ref.~\cite{elli1}.)  The shapes and
linear asymptotes of $r$ and $f^2$ are shown in Fig.~\ref{fig7}.  Not obvious,
but verifiable, is that $f^2(\rho) \sim 2 \, m/\rho$ as $\rho \to \infty$,
which, together with $r(\rho) \sim \rho$ as $\rho \to \infty$, shows $m$ to
correspond to the mass parameter of the Schwarzschild blackhole metric. 

The shared metric of the cross sections of constant $t$ is
$d\rho^2 + r^2 (\rho) \, d\Omega^2$.  The 2-sphere at $\rho = 2 \, m$ (where
$r(\rho)$ is a minimum) is the `threshold' of the drainhole `throat', crossed
when transiting from the `upper' region, where $\rho > 2m$, to the `lower'
region, where $\rho < 2m$, or vice versa.  Its areal radius is
$r(2 \, m) = n e^{(m/n) \alpha(2 \, m)}$, which increases monotonically from $n$
to $ne$ as $m$ increases from 0 to $n$.  Thus the size of the threshold is
determined primarily by $n$, and only ancillarily by $m$.  Moreover, as
indicated by the calculation that the contracted curvature tensor
$\text{\bf Ricci} = -2 (d\phi \otimes d\phi)
 = -\{n^2/[(\rho - m)^2 + a^2]^2\} (d\rho \otimes d\rho)$,
the strength of $\phi$'s `contribution' to the space-time geometry is
determined primarily by $n$ and is concentrated on the curvature of space,
providing the negative spatial curvatures necessary for the open throat to
exist.  In accord with the restriction adopted here that the `relaxant' scalar
fields $\phi$ and $\psi$ in the action integral of Eq.~(\ref{eqn6}) were not to
be varied in deriving the field equations, one would not say that $\phi$
{\it causes} (i.~e., is a source of) these spatial curvatures, but should
instead say that $\phi$ {\it allows} such negative curvatures to exist and
describes their configuration.  This understanding helps disabuse one of the
notion that geometrically and topologically unexceptionable space-time
manifolds such as the drainhole are somehow a product of `exotic' matter just
because their Ricci tensors disrespect some `energy condition' that traces back
to Einstein's 1916 assumption that inertial-passive mass and active mass are the
same.

Even when $m = 0$, so that there is no gravity, the throat stays open, with
$r(\rho) = \sqrt{\rho^2 + n^2}$.\footnote{The nongravitating drainhole is
frequently referred to as the `Ellis wormhole'~\cite{abe,naas}, as is
occasionally the gravitating drainhole~\cite{daka}.}  It is not a great stretch
to surmise that, whereas the parameter $m$ specifies the active gravitational
mass of the (nonexotic) drainhole particle, the parameter $n$ specifies in some
way its inertial rest mass.  This speculation is supported by two
considerations: first, as shown in Ref.~\cite{elli1}, the total `energy' of
the scalar field $\phi$ lies in the interval from $n/2$ to $n \pi/2$, thus is
essentially proportional to $n$; second, it would seem likely that the bigger
the hole (thus the larger is $n$), the greater the force needed to make it
move.  A `higgsian' way of expressing this idea is to say that the drainhole
`acquires' (inertial) mass from the scalar field $\phi$.

Because $r(\rho) \geq n > 0$ and $f^2(\rho) < 1$, the drainhole space-time
manifold is geodesically complete and has no one-way event horizon, the throat
being therefore traversable by test particles and light in both directions.
The vector field \mbox{$\partial_t + c f(\rho) \partial_\rho$} generates radial
geodesics parametrized by proper time; with $f$ chosen as
$-\sqrt{1 - e^{-(2 \, m/n) \phi}}$ it is taken to be the velocity field of a
`gravitational ether' flowing from $\rho = \infty$ on the high side (the upper
region), down into the drainhole throat, across the threshold, out into the low
side (the lower region), and onward to $\rho = -\infty$.  The ether's radial
acceleration is $(c^2 f^2/2)'(\rho)$, which computes to
$d^2 \! \rho/d\tau^2 = -c^2 m/r^2(\rho)$ and therefore is strongest at the
threshold, where $r(\rho)$ is at its minimum.  Because the radial acceleration
is everywhere aimed in the direction of decreasing $\rho$, the drainhole
attracts test particles on the high side and repels them on the low side.
Moreover, as demonstrated in Ref.~\cite{elli1} by means of an isometry of the
manifold that exchanges the upper and lower regions with one another, whereas
the upper region is asymptotic as $\rho \to \infty$ to a Schwarzschild manifold
with (active gravitational) mass parameter $m$, the lower region is asymptotic
as $\rho \to -\infty$ to a Schwarzschild manifold with mass parameter
$\bar m = -m e^{m \pi/a}$, so the drainhole repels test particles more strongly
on the low side than it attracts them on the high side, in the ratio
$-\bar m/m = e^{m \pi/\sqrt{n^2 - m^2}}$.  It is this excess of negative,
repulsive mass over positive, attractive mass that qualifies drainholes as
candidates for explaining the accelerating expansion of the universe and doing
away with the cosmological constant.

As mentioned in the Introduction, one can imagine that instead of a not well
defined `gravitational ether' it is space itself that flows into and through
the drainhole.\footnote{This idea has been applied in Ref.~\cite{hali} to
the interpretation and visualization of blackholes, with interesting results.}
This substitution, which as noted in the Introduction is in accord with
Einstein's insight that the concepts of space and of a gravitational ether are
essentially interchangeable, should cause no alarm, for the very notion of an
expanding universe already ascribes to space the requisite plasticity.  There
is, however, an ambiguity in that taking $f$ to be
$\sqrt{1 - e^{-(2 \, m/n) \phi}}$ instead of $-\sqrt{1 - e^{-(2 \, m/n) \phi}}$
would have space (or the ether) appear to be flowing up through the drainhole
rather than down.  Insisting that the velocity of the flow be accelerative, not
decelerative, would by fiat resolve the ambiguity in favor of the downward flow.

The discovery of the drainhole manifolds arose, in my case, from a search for
a model for gravitating particles that, unlike a Schwarzschild space-time
manifold, would have no singularity.  Geodesic completeness and absence of
event horizons resulted naturally from that requirement and the utilization of
a minimally coupled scalar field to weaken the field equations.  As shown in
Ref.~\cite{elli1} a drainhole possesses all the geodesic properties a
Schwarzschild blackhole possesses that do not involve its horizon or its
singularity, having eliminated the horizon and replaced the singularity with a
topological passageway to another region of space.  Drainholes are able,
therefore, to reproduce all the externally discernible aspects of physical
blackholes (if such things exist) that Schwarzschild blackholes reproduce.
That their low sides have never been knowingly observed (but in principle could
be) is no more troubling than the impossibility of directly observing the
insides of Schwarzschild blackhole horizons from external vantage points.  For
these reasons drainholes are more satisfactory than Schwarzschild blackholes as
mathematical models of centers of gravitational attraction.

\subsection{Dark matter and dark `energy' from drainholes}
\label{sec6.2}

A physical center of attractive gravity modeled by a drainhole could
justifiably be called a `darkhole', inasmuch as (as shown in
Ref.~\cite{elli1}) it would capture photons (and other particles) that
venture too close, but, unlike a blackhole, must with few exceptions
eventually release them, either back to the attracting high side whence they
came or down through the drainhole and out into the repelling low side.  Thus
one can imagine that at active galactic centers will be found not supermassive
blackholes, but supermassive darkholes.  At the center of our galaxy, for
example, instead of a blackhole of Schwarzschild mass
$m \approx 4 \times 10^6 M_\odot \approx 5.91 \times 10^6$~km (in geometric
units), and of horizon area
$4 \pi (2 \, m)^2 \approx 1.75 \times 10^{15}$ km$^2$, there might reside a
drainhole the area $4 \pi r^2(2 \, m)$ of whose threshold sphere would lie
between $4 \pi n^2$ and $4 \pi (ne)^2$, with $n$ constrained only by having to
exceed $m$.  As measured by the bounds $n$ and $ne$ on the areal radius
$r(2 \, m)$, and the corresponding bounds $c^2 m/n^2 = c^2 (m/n)(1/n)$ and
$c^2 m/(ne)^2 = c^2 (m/n)(1/e^2 n)$ on the maximum radial acceleration
$c^2 m/r^2(2 \, m)$, with $m$ constrained by $0 \leq m < n$, such an object
could be of any size and could be weakly gravitating for its size ($m \ll n$),
strongly gravitating for its size ($m \approx n$), or anything in
between.\footnote{A recent paper explores the possibility of observationally
distinguishing between a Schwarzschild blackhole and a nongravitating `Ellis
wormhole' at a galactic center.~\cite{tshaya}}

There is, however, more to be said.  A central tenet of the general theory of
relativity is that every object that gravitates, no matter how large or how
small, manifests itself in (more fundamentally, {\it is a manifestation of\/}) a
departure of the geometry of space-time from flatness.  If such an object has
other, nongravitational properties, these must be either incorporated in or
additional to the underlying geometric structure.  Believing that the drainhole
model provides the best presently available description of a gravitating
particle's geometry, I adopt the hypothesis that every such {\it elementary\/}
gravitating object is at its core an actual physical drainhole --- these objects
to include not only elementary constituents of visible matter such as protons
and neutrons (or perhaps, more elementarily, quarks), but also the unseen
particles of `dark matter' whose existence is at present only inferential.
Moreover, I assume that visible matter and the primordial matter of my
cosmological model are one and the same, and that drainholes with no additional
properties constitute the continuously created matter of the model.

The pure, isolated drainhole described by Eqs.~(\ref{eqn59}--\ref{eqn63}) is an
`Einstein--Rosen bridge' connecting two otherwise disjoint
`subuniverses'~\cite{eiro}, each of which, if evolving, would by itself consume
for its description all the resources of a Robertson--Walker metric.
Nonisolated drainholes presumably could exist not only as `bridges' between our
subuniverse and another (or multiple others), but also as `tunnels' from one
place in our subuniverse to another, possibly quite distant from the first by
every route that doesn't pass through a tunnel.  Both types could contribute to
the positive and the negative mass densities~$\mu$ and~$\bar \mu$ in our
subuniverse, each bridge drainhole contributing to~$\mu$ or to~$\bar \mu$, but
not to both, each tunnel drainhole contributing both to~$\mu$ by way of its
gravitationally attractive entrance portal and to~$\bar \mu$ by way of its
gravitationally repulsive exit portal.  Tunnel drainholes are easy enough to
visualize in abundance as topological holes into which flowing space disappears,
only to reappear elsewhere in our subuniverse, in analogy with rivers that go
underground and surface somewhere downstream.  Having both portals located in
our subuniverse, tunnel drainholes would have properties we could take fully 
into account.  Bridge drainholes, on the other hand, with only one side in our
subuniverse would have properties dependent in part on circumstances in other
subuniverses, circumstances beyond our ken.  I assume, therefore, that tunnel
drainholes contribute to $\mu$ and to $\bar \mu$, and bridge drainholes
contribute to neither, which leaves our subuniverse gravitationally
self-contained, but possibly connected to other subuniverses of a `multiverse'
by way of flowless bridge drainholes (nongravitating `Ellis wormholes').

Lacking for the present a full mathematical description of these tunnel
drainholes, one can nevertheless proceed under the assumption that they exist
and are characterized by parameters $m$ and $n$ related as in an isolated
bridge drainhole.  Every particle of gravitating matter, whether in the
`primordial' (P-matter) category or the `continuously created' (C-matter)
category, is then at its core one of these tunnels, and it becomes a question
of relating $m$ and $n$ to the present-epoch densities $\mu_{{\text P},0}$,
$\bar \mu_{{\text P},0}$, $\mu_{{\text C},0}$, and $\bar \mu_{{\text C},0}$.
A way to do this is developed in Ref.~\cite{elli4} (Sec.~VI.B), and will not
be repeated here.

\section{Issues and Observations}
\label{sec7}

The conceptual basis of the model of the cosmos developed in the preceding
sections differs in many respects from that of the standard, concordance model.
These differences and their consequences, each stemming directly or indirectly
from disallowing Einstein's implicit assumption that inertial mass produces
gravity, bring up a number of issues, some of which have been discussed above
to greater or lesser extent, others of which have been passed over.  In this
section I identify and comment on several of them.  For ease of reference I
will call my model the PCDM model (P~for~primordial, C for continuously created,
D for drainhole, M for matter) and the concordance model the $\Lambda$CDM
model ($\Lambda$ for $\Lambda$, C for cold, D for dark, M for matter).

\subsection{Gravity and passive-inertial mass}
\label{sec7.1}

As noted at the beginning, Newton's law of action and reaction allows the
inference that the ratio of active gravitational mass to passive-inertial mass
is the same for all `bodies', that is, for all matter in bulk.  Because
Newton's theory describes gravity as well as it does, this conclusion must be at
least approximately correct.  Even if exact, however, it is only a statement
about a numerical ratio of quantities that {\it conceptually\/} have nothing to
do with one another: the generation of a gravitational field on the one hand,
the resistance to being accelerated by a field of any sort on the other.
I pointed out in Sec.~\ref{sec2} that Einstein's theory of gravity in the
vacuum has in it no concept of inertial mass.  Subsequently I suggested
(in Sec.~\ref{sec6.1}) that the concept might now have been brought into the
vacuum theory as the parameter $n$ of the drainhole model.  If so, then the
particulars of the drainhole model make it quite clear that inertial mass and
active gravitational mass are entirely independent concepts.  Specifically, as
shown in~\cite{elli1}, when the active mass parameter $m$ is 0 but $n \neq 0$,
test particles follow the geodesics of the spatial metric
$d\rho^2 + (\rho^2 + n^2) d\Omega^2$.  These geodesics bend around the
drainhole, so give the appearance that the test particles are acted on
gravitationally by the drainhole.  But a test particle follows such a geodesic
at constant velocity, which can in fact be 0, leaving the test particle
sitting at rest for all time wherever it happens to be --- clearly not under
any gravitational influence.  In this way the drainhole model delineates the
distinction between the curvature of space, on one hand, and gravity (the
`curvature of time') on the other. 

If passive-inertial mass is not a source of gravity, then as said in
Sec.~\ref{sec3} Einstein's `energy-tensor of matter'
$T^{\alpha \beta} \defeq (\rho + p/c^2) u^\alpha u^\beta - p g^{\alpha \beta}$
has no role to play in the field equations of gravity, so one loses the
alluring implication that
$0 = {T_\alpha{}^\beta}_{:\beta}
= ((\rho + p/c^2) u^\alpha u^\beta)_{:\beta} - p_{.\beta} g^{\alpha \beta}$,
which implication Einstein interpreted as saying that a consequence of his field
equations was that ``the equations of conservation of momentum and energy\ldots
hold good for the components of the total energy'', and cited as ``the strongest
reason for the choice'' of his equations~(\S 16 of Ref.~\cite{eins2}).  The
loss of this implication does not, of course, keep us from asserting that
${T_\alpha{}^\beta}_{:\beta} = 0$; it only requires that we look elsewhere for
a justification.  But one must recognize that for (Einstein's)
${T_\alpha{}^\beta}_{:\beta} = 0$, however arrived at, to be interpreted as a
proposition about conservation of momentum and energy $\rho$ must be the
density of inertial mass, not, as Einstein assumed, the density of active
gravitational mass.

All too frequently there appear in the popular press accounts of astronomers
`weighing' the universe or `weighing' a galaxy.  One can object, of course, to
the presumably intentional blurring of the distinction between inertial mass
and weight, but a larger objection is that such accounts leave the reader with
the false impression that determining an entity's active gravitational mass is
known to be equivalent to determining its passive-inertial mass (and therefore
its weight).  Only if one can prove the existence of a universal
proportionality between the two can one legitimately claim to have `weighed'
the universe or a galaxy, and even then only indirectly.  Unless and until
there is such a proof, pretending that the universe or a galaxy has been
weighed spreads ignorance, not knowledge.  The most elementary unit of
knowledge is a distinction made --- a fact every digital computer is based on.
Conversely, the most elementary unit of ignorance is a distinction not made.

\subsection{Choice of a variational principle for gravity}
\label{sec7.2}

The step up in Sec.~\ref{sec3} from the nonrelativistic variational principle
$\delta \! \bigintsss \! (|\nabla \phi|^2 + 
                         8 \pi \kappa \mu \phi) \, d^3\!x = 0$
that generates the Poisson equation $\nabla^2 \phi = 4 \pi \kappa \mu$ to the
relativistic principle
$\delta \! \bigintss \! ({\bm R} - \textstyle\frac{8 \pi \kappa}{c^2} \mu)
                     \, |g|^{\frac12} d^4\!x = 0$
that generates the equations
${\bm R}_{\alpha \beta} - \textstyle{\frac12} {\bm R} \, g_{\alpha \beta}
 = -\frac{4 \pi \kappa}{c^2} \mu g_{\alpha \beta}$
is an application of Occam's razor, and as such can only be justified
retrospectively, by its consequences.  Alone, this principle has, through its
Euler-Lagrange equations, little to say about cosmology, but when modified by
inclusion of the cosmological constant $\Lambda$ it says something significant,
namely, that there is gravitationally repulsive matter in the universe
(disguised as $-\Lambda$), and there can be more of it than there is of
gravitationally attractive matter.  This is a consequence that has consequences,
but to get them all requires the next step up, to the action integrand of
Eq.~(\ref{eqn6}) which has in it in addition to the positive and negative mass
densities $\mu$ and $\bar \mu$ the (gradients of the) scalar fields $\phi$
and~$\psi$, which are considered as auxiliary `relaxants' not to be varied.
What justifies their inclusion?

A defect of the Einstein vacuum field equations, generally unrecognized as such,
is that they produce as their basic model for a gravitating particle a manifold
that exhibits no {\it spatial} curvature when viewed from the perspective of
observers free-falling from rest at infinity.  (This is seen in
Eq.~(\ref{eqn59}), in which $t$ coincides with the proper time of such observers
and on $t = {\rm constant}$ cross sections of the Schwarzschild manifold the
metric is that of (flat) euclidean 3-space, viz., $d\rho^2 + \rho^2 d\Omega^2$.)
Inclusion of the scalar field $\phi$ in the action integral of Eq.~(\ref{eqn6})
corrects this defect, as it allows the constant-$t$ cross sections of the
drainhole model to have the negative spatial curvatures characteristic of the
drainhole, and therefore admits the possibility that the space we reside in is
something other than euclidean 3-space.

The inclusion of a second scalar field $\psi$, coupled to the metric with
polarity opposite to that of $\phi$'s coupling, is dictated in the first
instance by the absence, when Einstein's assumption of equivalence between
passive-inertial mass and active gravitational mass is denied, of any real
reason to choose one polarity over the other.  In the second instance it is
dictated by the observation that, although the presence of $\phi$ but not of
$\psi$ was required for the derivation of the drainhole model, in the
construction of the cosmological model absence of $\psi$ would entail absence
from Eq.~(\ref{eqn22}) of the $\dot \beta^2$ term, in which case
Eq.~(\ref{eqn22}) would turn from true to false once $R(t)$ surpassed
$\sqrt[3]{B/A_{\text P}}$ (which is $R_{H_{\text{max}}}$ when $k = 0$).
Moreover, if the shoe were on the other foot and the $\dot \alpha^2$ term were
absent, then Eq.~(\ref{eqn22}) would be false when $R(t)$ was less than
$\sqrt[3]{B/A_{\text P}}$.  Thus both $\phi$ and $\psi$ are needed, coupled to
the geometry with opposite polarities. One notices that $\alpha$ and $\beta$,
appearing only in the combination $\dot \alpha^2 - \dot \beta^2$, can be
individuated only by an arbitrary allocation of the righthand side of
Eq.~(\ref{eqn22}) that preserves positives and negatives, such as
$\dot \alpha^2 = c^2 B/R^6 + h$ and
$\dot \beta^2 = c^2 A_{\text P} R^3/R^6 + h$, where $h$ is a nonnegative
function of~$t$.

That the combination
$2 \, \phi^{.\gamma} \phi_{.\gamma} - 2 \, \psi^{.\gamma} \psi_{.\gamma}$
in the action integrand of Eq.~(\ref{eqn6}) is the real part of
$2 \, \chi^{.\gamma} \chi_{.\gamma}$, where $\chi \defeq \phi + i \psi$,
suggests that space-time as seen here might be a restriction of a more general,
at least partly complexified space-time geometry.

\subsection{Inflation and the `big bounce'}
\label{sec7.3}

Further evidence that generation of a realistic cosmological model depends on
inclusion of the scalar field $\phi$ with the so-called `ghost' or `phantom'
coupling to geometry comes from the recognition that without it there is a
`bang' singularity at $R_{\text{min}} = 0$ and no inflation to follow, whereas
with it there is a nonsingular bounce at $R_{\text{min}} > 0$, followed by
inflation.  As is evident from the definition of $P_1(R)$ in Eq.~(\ref{eqn25}),
as well as from examining Fig.~\ref{fig1}, for $R_{\text{min}}$ to be positive
it is necessary and sufficient that the integration constant $B$ be positive
(otherwise $P_1(0) = -B \geq 0$, which would make $R_{\text{min}} = 0$).  Also
evident is that the smaller a positive $B$ is, the closer $R_{\text{min}}$ is
to~0; in fact, as noted in Sec.~\ref{sec4},
$R_{\text{min}} \sim \sqrt[3]{B/2 A_{\text P}}$ as $B \to 0$.  Moreover, it is
clear that the term $2 B/3 R^6$ in Eq.~(\ref{eqn24}) is the term that produces
rapid inflation when $R$ is small by making $\ddot R$ large, but only when
$B > 0$.  And because
$B - A_{\text P} R_{\text{min}}^3 \sim B - A_{\text P} (B/2 A_{\text P}) = B/2$
as $B \to 0$, positivity of $B$ demands that the ${\dot \alpha}^2$ term be
present in Eq.~(\ref{eqn22}), thus that the scalar field $\phi$ be present in
the field equations, coupled to the space-time geometry with the unconventional
polarity.  Let us note, moreover, that here there is no mention of a `slow roll
down a potential', a common element in many proposed inflationary scenarios
involving an `inflaton' scalar field.  Indeed, that $\phi$ is treated merely as
an aid in describing the space-time geometry rather than as a `physical source'
of the geometry makes the notion of introducing a potential function for $\phi$
irrelevant.

\subsection{Dark matter, dark `energy', and the `Cosmological Constant
Problem'}
\label{sec7.4}

In the $\Lambda$CDM model dark matter is lumped with baryonic matter in
$\Omega_{\text M}$ of the Friedmann equation (Eq.~(\ref{eqn58})) as primordial
matter whose density is proportional to $1/R^3$.  The unseen, unknown stuff in
$\Omega_\Lambda$, whose density has no $R$ dependence, is called dark `energy'
in compliance with Einstein's assumption that inertial mass, therefore energy,
is a source of gravity.  In the PCDM model only baryonic matter is treated as
belonging to the primordial $\Omega_{\text P}$ sector with $1/R^3$ density
dependence, dark matter being located in the $\Omega_{\text C}$ sector with
density kept constant by continuous creation of the dark \text{C-matter}.  And
there, instead of dark `energy' represented by a cosmological constant $\Lambda$
required by the data to be many orders of magnitude too small to be consistent
with the `vacuum energy' proposed as its source, one finds the gravitationally
attractive dark matter particles accompanied by their gravitationally repulsive
back sides, their net effect represented by the constant $A_{\text C}$ of
Eq.~(\ref{eqn21}), functionally equivalent to $\Lambda$ and with the requisite
smallness readily producible by drainholes.\footnote{For the best fit value
0.954 of $\Omega_{\text C}$ mentioned in Sec.~\ref{sec5.2},
$A_{\text C} = (3 H_0^2/c^2)\Omega_{\text C} = 1.821 \times 10^{-52}$m$^{-2}$.
Sec. VI.B of Ref.~\cite{elli4} shows how drainholes can produce such a
value.} This interpretation of `dark energy' as the back side of dark matter
provides a self-consistent solution of the so-called `Cosmological Constant
Problem'.

\subsection{Ratio of dark matter to baryonic matter}
\label{sec7.5}

Because in the $\Lambda$CDM model the densities of dark matter and baryonic
matter are both proportional to $1/R^3$, the ratio of their densities is
constant in time.  Estimates of that ratio obtained from observations of
visible structures such as galaxies and galactic clusters and superclusters as
they exist at the present epoch in the vicinity of our galaxy are therefore
used in studies of the formation of these structures in the distant past.  In
the PCDM model, on the other hand, the dark matter density is held constant
through continuous creation of \text{C-particles}, so the ratio
$\mu_{\text C}/\mu_{\text P}$ of dark \text{C-matter} density to baryonic
\text{P-matter} density grows in proportion to $R^3$.  Consequently, the value
of this ratio at the epoch corresponding to redshift $z$ is
$\gamma(z) = \gamma_0/(1 + z)^3$, where $\gamma_0$ is the ratio at present.
Taking this dependence of the ratio on redshift as real would no doubt
influence the outcomes of studies of structure formation.

\subsection{Continuous creation of \text{C-matter} tunnels}
\label{sec7.6}

The stipulation that the net density $\mu_{\text C} + \bar \mu_{\text C}$ of
\text{C-matter} stay fixed while the universe is expanding requires that the
drainhole tunnels considered to be the particles of \text{C-matter} come
continuously into existence.  By what mechanism might this happen?  Wheeler's
notion of a `quantum foam' of wormholes popping into and out of existence would
not provide a satisfactory explanation, for in order to keep the density
constant each newly created tunnel must either remain in existence or else upon
dying be multiply replaced by new ones.  A more useful idea is that the tunnels
arise from the stretching of space as the universe expands.  Once started, this
would be a self-sustaining process, each new tunnel causing by its excess of
repulsion over attraction additional stretching that would generate additional
tunnels.  Such a process, moreover, could be expected to produce tunnels in a
given region of space at a rate proportional to the rate of increase of the
volume of that region, thus maintaining a constant tunnel particle density.  Not
only that, these tunnels, once in existence, might very well themselves expand
along with the stretching of space, which they could do without altering their
active gravitational masses, thus without changing the net density.  As to how
the process could begin, one looks to the primordial \text{P-matter} tunnels,
presumed to have always been present, pushing the universe toward expansion.
In the postbounce era the \text{C-particles} would add their contribution, the
ultimate result being an acceleration of the expansion.  In the prebounce era
the \text{C-particles}, abundant in the distant past, would be dying out with
the shrinking of space as the universe contracted into the bounce.\footnote{The
once seemingly moribund idea of a universe held in a `steady state' by
continuous creation of matter through the agency of a `negative-energy' scalar
field\cite{vina,napa} has been to a considerable extent reinvigorated by the
recognition that drainholes and other such `traversable wormholes' demand the
`wrong coupling' for their existence.}

If the primordial \text{P-matter} has `always' existed, then the question of
what `caused' it to exist is better left to the philosophers.  Conceivably,
however, the \text{P-matter} tunnels came into existence as dark matter
particles the same way the \text{C-matter} tunnels did, but in the inflationary
space-stretching era after the bounce.  Only later would their entrance portals
have undergone modifications that turned them into particles of visible matter.
The really fundamental unanswered question, though, is how did space and time
come to exist.  In fact, the question can be reduced to the existence of space
alone, for the Kaluza--Weyl theory described in Ref.~\cite{elli5} (and more
succinctly in Ref.~\cite{elli6}) is based on a construction that produces
time (thus space-time) from three-dimensional space and, repeated, produces a
secondary time (thus space-time--time) from space-time.  And at the space-time
stage of that theory there occurs naturally a model of an expanding drainhole
in an expanding universe which repels on the low side more strongly than it
attracts on the high side and needs no auxiliary scalar field for its
description~\cite{elli7}.

It cannot be ruled out that some \text{C-particles} (and some
\text{P-particles} as well) have active gravitational mass $m = 0$, but,
because $n \neq 0$, have nonzero inertial mass.  Being without electric charge,
and possessing some rotational properties as suggested in
Ref.~\cite{manu} to be possible, such tunnel particles might be produced in
abundance and could serve as models for the ubiquitous cosmic neutrinos.  If
such models are realistic, then analyses that use cosmological observations to
put upper bounds on the neutrino mass scale (the analysis in
Ref.~\cite{thabla} using observations of galaxy clustering, for example) say
nothing about the inertial masses of neutrinos, instead `constrain' only their
active gravitational masses, already assumed in the models to be zero.  Even if
these models are not realistic, in the absence of any demonstrable relation for
neutrinos between the two kinds of mass such analyses based on gravitational
effects cannot be presumed to constrain their inertial masses.

One other possibility to consider is that as time goes on a fraction of the
\text{C-particles} acquire the trappings of ordinary matter and are then able
to join in the formation of visible-matter structures.  This would offer a way
for the universe to avoid a cold, dark ending to its exponentially accelerating
expansion.

\subsection{Voids, walls, filaments, and nodes}
\label{sec7.7}

The PCDM model, just as the $\Lambda$CDM model, treats matter as uniformly
distributed in space, an apparently reasonable assumption on a large enough
scale.  Observationally, however, a major fraction of space appears to comprise
voids nearly empty of visible matter, separated by walls, filaments, and nodes
in which resides most of the matter, both the dark and the visible.  The
formation of this cellular structure is generally considered to be an
evolutionary product of early small fluctuations in the density of primordial
matter.  Explanations along that line are attempts to realize what Peebles has
called his ``perhaps desperate idea \ldots that the voids have been emptied by
the growth of holes in the mass distribution''~\cite{peeb}.  If besides matter
that attracts gravitationally there is also matter that repels, then that idea
can be reinterpreted and the desperation perhaps alleviated, by the simple
observation that, whereas attractive matter wants to congregate, repulsive
matter is reclusive, pushing all matter, attractive or repulsive,
away~from~itself.

In the PCDM model each tunnel particle would presumably be created with its
entrance and its exit close to one another in the ambient space.  In the
ambient space the entrance would attract the exit, but the exit would repel the
entrance more strongly, so the two portals would drift apart.  Apply this to a
multitude of such particles and you might expect to see the exits spread
themselves over regions from which they had expelled the entrances, regions
therefore devoid of attractive matter.  The entrances, on the other hand, being
brought together by both their mutual attractions and the repulsion from the
exits, would aggregate into walls, filaments, and nodes on the boundaries
between the void regions, just as is seen in the real universe.  What is more,
the walls, filaments, and nodes so created would likely be, in agreement with
observation, more compacted than they would have been if formed by gravitational
attraction alone, for the repulsive matter in the voids would increase the
compaction by pushing in on the clumps of attractive matter from many directions
with a nonkinetic, positive pressure {\it produced by\/} repulsive gravity, a
pressure not to be confused with the negative pseudo-pressure conjectured in the
confines of Einstein's assumption to be a {\it producer of\/} repulsive gravity.
In this way the PCDM model can explain qualitatively the observed cellular
structure of the universe with minimal reliance on preexisting fluctuations in
the primordial matter density --- it is conceivable that creation of tunnel
entrance-exit pairs at random locations in random orientations during the
inflationary phase would provide in and of itself all the density fluctuations
needed to initiate the formation of the voids, walls, filaments, and nodes as
they are seen today.

\subsection{Protons, neutrons, and WIMPs as drainholes}
\label{sec7.8}

The cosmological model developed in Secs.~\ref{sec4} and \ref{sec5} and
fitted to the SNe Ia data rested on the assumption that the active
gravitational mass densities of primordial matter and continuously created
matter are on balance negative.  The model did not include (or need to include)
a mechanism for producing those imbalances.  Subsequently, in
Sec.~\ref{sec6.2}, I proposed as a mechanism that every elementary particle
that gravitates is at its core a physical drainhole tunnel, and that the
excesses of repulsion over attraction of these tunnels cumulatively produce the
overall density imbalances.  Among the particles in that category would
presumably be counted protons, neutrons, and the weakly interacting massive
particles (WIMPs) thought to be the constituents of dark matter.  From my
perspective the drainhole proposal is simply to modify the conception of every
such particle as an entity built around a physical blackhole, by substituting
for the blackhole a physical drainhole tunnel.  Such a substitution seems to me
no more radical than replacing a blocked drainpipe of a wash basin with a
drainpipe that has no blockage.  But one must in the first place recognize that
the basin {\it has} a drainpipe, whether blocked or unblocked.  I find it more
than a little peculiar that there is discussion about the possibility of the
Large Hadron Collider's producing a microscopic blackhole through a collision of
two protons, when a straightforward application of Einstein's unvarnished theory
of gravity suggests that every proton and every neutron in the collider is
already built around a blackhole or a collection of blackholes.  Are not
gravitating bodies just conglomerates of gravitating particles, each making its
own small contribution to the gravity of the whole?  And if a particle that
gravitates has no gravitational sink such as a blackhole or a drainhole
associated with it, then what {\it is} the source of its gravity?

\subsection{Galactic nuclei as drainholes}
\label{sec7.9}

If instead of a blackhole at the center of our galaxy there is a drainhole,
what might it look like?  From afar the behavior of matter and radiation at
some distance outside such a drainhole would differ very little in appearance
from the behavior of matter and radiation at the same distance from the event
horizon of a blackhole of the same active mass.  Closer in, absence of a
horizon in the drainhole makes a big difference.  For both a Schwarzschild
blackhole of mass $m$ and a drainhole of mass $m$ the radial equation of motion
of a test particle is
\vskip -10pt
\begin{equation}
\frac{d \hspace{0.5pt}^2 \! \rho}{d\tau^2}
 = \displaystyle -\frac{c^2 m}{r^2(\rho)}
                  + (\rho - 3m) \! \left(\frac{d \Omega}{d\tau}
                                   \right)^{\!\! 2} \, ,
\label{eqn64}
\end{equation}
with $r(\rho)$ given by Eq.~(\ref{eqn62}) for the drainhole and
$r(\rho) = \rho$ for the blackhole.  In both cases $\tau$ is the proper time of
the test particle.  Equation~(\ref{eqn64}) shows that for the drainhole as well
as for the blackhole $3m$ is the radius of its `capture sphere', as any test
particle whose orbit takes it from above $\rho = 3m$ to $\rho = 3m$ or below
will never again have $d\rho/d\tau > 0$, and likewise for photons, $\tau$ being
in that case any affine parameter and the first term on the right being absent.
Note, however, that upward moving test particles can with sufficient radial
velocity rise above $\rho = 3m$ and remain there, and that upward moving photons
with angular velocity not too large can escape to $\rho = \infty$, so luminous
objects falling into the drainhole and out the far side can in principle be
watched from above forever.\footnote{A complete descriptive catalog of the
geodesics of drainholes can be found in Ref.~\cite{elli1}.} 

The metric of every cross section of constant $t$ being
$d\rho^2 + r^2 (\rho) \, d\Omega^2$, the cross sections of the blackhole are
euclidean, so the geodesic radius $3m$ of the capture sphere is also its areal
radius.  For the drainhole the situation is different: the areal radii $r(2m)$
of the threshold sphere and $r(3m)$ of the capture sphere can be made as large
as you like by increasing the size parameter $n$, while leaving the mass
parameter $m$ unchanged, inasmuch as $m < n < r(2m) < r(3m)$, whatever values
$m$ and $n$ have.

An analysis in~\cite{elli4} of tidal gradients in the drainhole finds that an
object free-falling downward encounters a maximum stretching at
$\rho = 2m + n/\sqrt{3}$ (above the hole) and a maximum compression
(antistretching) at $\rho = 2m - n/\sqrt{3}$ (below the hole).  For a fixed $m$
these have limits proportional to $c^2/m^2$ as $n/m \to 1$, and decrease
monotonically to 0 as $n/m \to \infty$.  With
$m = m_{\text{gc}} \approx 4.31 \times 10^6 M_\odot = 6.36 \times 10^6$~km
in geometric units (the current best estimate of the supermass at the center of
our galaxy~\cite{gill}), a star the size of our Sun falling radially downward
would experience a maximum stretching difference of accelerations between the
leading edge and the trailing edge of 56 km~s$^{-2}$ if $n = 2 m_{\text{gc}}$
and a difference of 8 km~s$^{-2}$ if $n = 5 m_{\text{gc}}$.  The corresponding
numbers for the maximum compressing differences are 115 km~s$^{-2}$ and
11 km~s$^{-2}$.  These are small enough that the star would likely survive the
journey and show up intact in the region the drainhole exits to on its downside.
The same would be true for gas and dust clouds.  This is in stark contrast to
the supermassive blackhole scenario, in which every bit of matter that falls in
encounters unbounded tidal stretching that strips it of its identity as it
approaches and reaches the singularity at $r = 0$, never to be seen again.

If supermassive drainholes at galactic centers are commonplace, and their exits
reside in our (part of the) universe, where should we expect to see those exits,
and how might we identify them?  In accordance with the notion that the exits of
drainhole tunnels tend to isolate themselves by pushing all other matter away,
the obvious places to search would be the great voids created by the exits of
both the \text{P-matter} and the \text{C-matter} tunnels.  Presumably they would
appear as objects of the same sizes as their entrances, emitting pure light
intrinsically blueshifted from its journey downward in the drainholes'
gravitational potentials, as well as clouds of outflowing gas (and perhaps
fleeing stars).  The brightnesses of such objects would depend on the rate at
which the gases and light were sucked into the drainhole entrances.  There being
no reports of any such bright objects observed in the voids, one would appear to
be left with the alternative that they are there but are too dim to be seen.
There is, however, another possibility that seems to me at least as likely.

If, as I proposed in Sec.~\ref{sec6.2}, every elementary particle that
gravitates is at its core a physical drainhole tunnel, then gravitationally
every gas cloud or star is simply a conglomeration of entrances of tiny tunnels
whose exits are spread throughout the bubble voids of our universe.  Each of
these little tunnels would be taking in a commensurably small portion of the
gravitational ether and pushing it out in a void, there to do its part in
causing the bubble (and therefore the universe) to expand.  A neutron star would
be gravitationally a sieve-like, very dense collection of tiny tunnel entrances;
supernova cores would collapse gravitationally to superdense such collections
rather than to blackholes.  At the center of our Milky Way galaxy there could be
then, not a supermassive drainhole, but a supermassive star, formed from the
melding of multitudes of gas clouds and stars brought together by their mutual
gravitational attractions, comprising gravitationally $4.31 \times 10^6$ as many
of those tiny drainhole tunnel entrances as are in our Sun --- a `supersun' or
`superstar', so to speak.

There is yet another possibility worth considering.  With a little topological
imagination one can see how a supermassive drainhole might arise from a
plenitude of small dark \text{C-matter} drainhole tunnels with their entrances
packed
tightly together by their mutual attractions.  Begin with a pair of them, $T_1$
and $T_2$ say.  If not only their entrances but also their exits were close
together, then the coalescing of $T_1$ and $T_2$ into a single tunnel might be
feasible.  If, however, their exits had receded far from one another, then the
only way $T_1$ and $T_2$ might reasonably coalesce into one tunnel would be to
join at their entrances and close off into a single tunnel connecting their
original exits.  If that occurred, then the `gravitational ether' flowing into
their entrances would be diverted to the nearby entrances of one or more
tunnels, thereby increasing their masses, both the attractive and the repulsive.
The remnant of $T_1$ and $T_2$, no longer connected to the ether flow, would
become a gravityless tunnel connecting two distant places out in the voids
(perhaps, as suggested in Sec.~7.6, manifesting as a neutrino --- or a pair of
neutrinos).  After many repetitions of this process there would be left a single
tunnel large enough to accommodate all the combined ether flow of the
\text{C-matter} tunnels --- a supermassive, dark drainhole.  To account for the
exit of such a drainhole's being too dim to be observed in the voids, one would
have to presume that little if any of the matter in its accretion disk
`accretes' down the drainhole, the majority of it instead being converted into
polar jets by one or another of the mechanisms proposed for generation of
blackhole jets~\cite{abfr}, adapted for drainholes.

\section{Summary}
\label{sec8}

Exploring the consequences of denying Einstein's 1916 assumption that inertial
mass and energy are sources of gravity brings one to the variational principle
\begin{equation}
\delta \! \int [{\bm R} - \textstyle{\frac{8 \pi \kappa}{c^2}} (\mu + \bar \mu)                         + 2 \, \phi^{.\gamma} \phi_{.\gamma}
                        - 2 \, \psi^{.\gamma} \psi_{.\gamma}] \,
               |g|^{\frac12} d^4\!x = 0,
\nonumber
\end{equation}
in which $\phi$ and $\psi$ are scalar fields and $\mu$ and $\bar \mu$ are the
{\it active\/} gravitational mass densities of distributions of gravitationally
attractive and gravitationally repulsive matter, and in which, to accord with
the precept that in a space-time manifold nothing extraneous to the metric
should participate in the extremizing of the action, only the space-time metric
is varied in deriving the field equations to govern the space-time geometry.
This logically consistent, purely geometrical version of Einstein's theory of
gravity is then found to be capable of performing feats the original version
could not perform.  Specifically, the modified version is found to be able to:

\begin{itemize}

\item Produce cosmological models that replace the `big bang' with a `big
bounce', include in their expansion inflation, deceleration, coasting, and
ultimate exponential acceleration, and provide good fits to Hubble plots of
type~Ia supernovae data.

\item Solve the `Cosmological Constant Problem', by identifying
$-\frac{c^2}{4 \pi \kappa} \Lambda$ as the net active mass density of
gravitating matter.

\item Replace the Schwarzschild blackhole with a singularity-free, horizonless,
topological `drainhole' that gravitationally attracts matter on its high, front
side while gravitationally repelling matter more forcefully on its low, back
side.

\item Represent dark matter by the attracting entrance portals of drainhole
tunnels and dark `energy' by the repelling exit portals of those tunnels.

\item Suggest a new, drainhole-based mechanism for the creation and evolution
of cosmic voids, walls, filaments, and nodes.

\item Exorcise `phantoms', `ghosts', and `exotic matter' from the body of
gravitational physics.

\end{itemize}

\noindent
In view of these successes, as well as its logical consistency, one can
justifiably consider Einstein's theory of gravity modified in this way to be an
improved version of the original.

\end{document}